# Flow similarity, stochastic branching, and quarter power scaling in plants

Charles A. Price[1,2], Paul Drake[2,3], Erik J. Veneklaas[2,3] and Michael Renton[2,3]

[1]National Institute for Mathematical and Biological Synthesis (NIMBioS)
University of Tennessee
Knoxville, TN, 37996-3140
USA

[2]School of Plant Biology
University of Western Australia
Perth, Western Australia 6009
Australia

[3]Centre of Excellence for Climate Change, Woodland and Forest Health
University of Western Australia
Perth, Western Australia 6009
Australia

**Corresponding Author**:
Charles A. Price
National Institute for Mathematical and Biological Synthesis
University of Tennessee
Knoxville, TN, 37996-3140
USA
Phone: (865) 974-9334
Fax: (865) 974-9300
Email: chuckprice69@gmail.com

**Summary**

- The origin of allometric scaling patterns that are multiples of ¼ has long fascinated biologists. While not universal, scaling relationships with exponents that are close to multiples of ¼ are common and have been described in all major clades. Foremost among these relationships is the ¾ scaling of metabolism with mass which underpins the ¼ power dependence of biological rates and times.

- Several models have been advanced to explain the underlying mechanistic drivers of such patterns, but questions regarding a disconnect between model structures and empirical data have limited their widespread acceptance. Notable among these is a fractal branching model which predicts power law scaling of both metabolism and physical dimensions. While a power law is a useful first approximation to many datasets, non-linearity in some large data compilations suggest the possibility of more complex or alternative mechanisms.

- Here, we first show that quarter power scaling can be derived using only the preservation of volume flow rate and velocity as model constraints. Applying our model to the specific case of land plants, we show that incorporating biomechanical principles and allowing different parts of plant branching networks to be optimized to serve different functions predicts non-linearity in allometric relationships, and helps explain why interspecific scaling exponents covary along a fractal continuum. We also demonstrate that while branching may be a stochastic process, due to the conservation of volume, data may still be consistent with the expectations for a fractal network when one examines subtrees within a tree.

- Data from numerous sources at the level of plant shoots, stems, petioles, and leaves show strong agreement with our model predictions. This novel theoretical framework provides an easily testable alternative to current general models of plant metabolic allometry.

**Introduction**

Since Max Kleiber first examined the scaling of animal metabolism with mass (Kleiber 1932), scientists have been interested as to why allometric relationships often have exponents that are close to a multiple of ¼ (Brown & West 2000). Following the early works of Brody (1945) and Hemmingsen (1950), several seminal books published in the 1980's expanded the number and breadth of relationships that exhibit approximate quarter power scaling, further generating interest in this area (McMahon & Bonner 1983; Peters 1983; Calder 1984; Schmidt-Nielsen 1984) and helping to establish a "mystery of quarter power scaling in biology" (Brown & West 2000).

The publication of West, Brown and Enquist's (WBE) fractal branching model (West, Brown & Enquist 1997; West, Brown & Enquist 1999b), which proposes a mechanism to explain the origin of quarter power scaling relationships, further catalysed interest in this area. WBE argued that a scaling relationship between organism volume, and the surface area available for resource exchange, should ultimately drive quarter power scaling, and suggested that while external surface area in mammals could follow a geometric scaling (Rubner 1883), internal vessel network geometry might be fractal, yielding quarter power scaling and effectively giving life a "fourth dimension" (West, Brown & Enquist 1999a). Subsequent efforts to derive optimal network geometries invoke supply/demand arguments (Banavar, Maritan & Rinaldo 1999; Banavar *et al.* 2014), or volume minimization (Dodds 2010). A common feature of these approaches is that they search for global optima and assume that fluid loss occurs at distributed sinks which are typically modelled as the ends of vessels. However, in both plants and animals, different parts of the fluid distribution network may be optimized to perform different functions (Murray 1926; Price, Knox & Brodribb 2013), and fluid is usually lost transmurally (Zwieniecki *et al.* 2002): vessel endpoints are not the usual mode of fluid exchange.

While the WBE model has been invaluable in helping to generate interest in biological scaling, unanswered questions regarding the disconnect between empirical data, model assumptions, and predictions have limited its widespread acceptance (Dodds, Rothman & Weitz 2001; Niklas 2004; Coomes 2006; Price, Enquist & Savage 2007; Savage, Deeds & Fontana 2008; Price *et al.* 2012). Several reports have indeed shown that proxies for metabolic rate in mature trees do scale with exponents close to the predicted ¾ (Niklas & Enquist 2001; Meinzer *et al.* 2005; Mori *et al.* 2010), but as recently highlighted by Price et al. (2012), empirical exponents by themselves do little to help determine an underlying mechanism. Given empirical support for a ¾ scaling of metabolism, arguably the strongest subsequent test of WBE is whether or not the geometry of biological distribution networks conforms to the specific fractal structure that is invoked. Results from several studies suggest that while branching is consistent with the assumed area-preserving architecture, the scaling of branch lengths is largely inconsistent with the WBE "volume-filling" assumption (Price, Wing & Weitz 2011; Bentley *et al.* 2013; Tredennick, Bentley & Hanan 2013).

Here, we suggest that plant distribution networks may indeed have "fractal-like" characteristics, but that these characteristics differ in important ways from those described by WBE. We show that a network which conserves volume flow rate and velocity, which we refer to collectively as "flow similarity", also exhibits a ¾ scaling relationship between surface area and volume. Subsequent incorporation of size-dependent hydraulic and biomechanical constraints leads to non-linear predictions for numerous allometric patterns. To test this novel theory for the scaling of plant architecture and metabolism, we analyse the geometry of vascular plant networks at four scales of organization: 1) whole plant shoots, 2) terminal stems, 3) petioles, and 4) leaf veins.

**Symmetric branching**

We begin by considering the hydraulic behaviour of the terminal branches in plants. Terminal branches in tracheophytes generally, and in woody species in particular, are where the overwhelming majority of leaves are borne and thus they constitute the predominant sites of photosynthesis and production. We follow previous work (Shinozaki *et al.* 1964; West, Brown & Enquist 1997; West, Brown & Enquist 1999b; Savage *et al.* 2010) in modelling branching as an idealized, symmetric, branching flow network. For now we assume that locally, the number of internal conduits scales linearly with the number of external branches, $n_{int} \propto n_{ext}^p$, where p ≈ 1. Under such an assumption, the scaling of the internal vessels parallels that of the external branches thus, for the purposes of the following derivation we don't differentiate between the two. We consider exceptions to this assumption in the Discussion and Supplementary Note 1.

Under symmetric branching, the ratio of the daughter (*k+1*) to parent (*k*) branch radii is $r_{k+1}/r_k = n^{-a}$, where *n* is the number of daughter branches. If we assume that flow velocity is constant across branching generations, we have area preserving branching where *a*=1/2, and thus

$$r_{k+1} = n^{-1/2} r_k \qquad (1)$$

Volumetric flow rate (*Q*) through a conduit within the network can be approximated via the well known Hagen-Poiseuille equation as $Q = \frac{\pi r^4 |\Delta P|}{8 \eta l}$, where *r* is conduit radius, *l* is conduit length, Δ*P* is the difference in pressure between the ends of the conduit, and *η* is viscosity. If we assume that $\eta$ and $\Delta P$ are locally constant (the same in parent and daughter branches, see Supplemental Note 1), then we have

$$l_k = \frac{c r_k^4}{Q_k} \quad (2)$$

and

$$l_{k+1} = \frac{c r_{k+1}^4}{Q_{k+1}} \quad (3)$$

where $c = \frac{\pi |\Delta P|}{8\eta}$ and thus (from Eq. 2) the ratio of radius squared to length in the parent branch is

$$\frac{r_k^2}{l_k} = r_k^2 \frac{Q_k}{c r_k^4} = \frac{Q_k}{c r_k^2} \quad (4)$$

Furthermore, with symmetry the volumetric flow from the parent branch is divided evenly among its daughters, and so

$$Q_{k+1} = Q_k / n = n^{-1} Q_k \quad (5)$$

Based on Eq. 4, the ratio of radius squared to length in the daughter branch is

$$\frac{r_{k+1}^2}{l_{k+1}} = \frac{Q_{k+1} r_{k+1}^2}{c r_{k+1}^4} = \frac{Q_{k+1}}{c r_{k+1}^2}$$

$$= \frac{n^{-1} Q_k}{C(n^{-1/2} r_k)^2} \quad \text{(based on Eqs 1\&5)}$$

$$= \frac{n^{-1} Q_k}{C n^{-1} r_k^2} = \frac{Q_k}{C r_k^2}$$

$$= \frac{r_k^2}{l_k} \quad (6)$$

This means that the ratio of radius squared to length is the same in the daughter and parent branch, and since this can be shown for any daughter/parent branch combination within the local structure, we have the general relationship

$$l \propto r^2 \qquad (7).$$

Using standard formulas for the surface area, $SA = 2\pi r l$ and volume, $V = \pi r^2 l$, of a cylinder, together with Eq. 7, we have $SA \propto 2\pi r^3$ and $V \propto \pi r^4$, and so

$$SA \propto V^{3/4} \qquad (8).$$

These scaling arguments apply to the individual internodes within a tree, but in Supplementary Note 2 we show that the $l \propto r^2$ and $SA \propto V^{3/4}$ scaling result can be extended to subtrees, and indeed the entire tree. Additional relationships between the length, diameter, surface area and volume in fractal trees follow easily from Eqs. 7 and 8 (Table 1). Thus, only two physical principles, the conservation of volumetric flow rate and velocity across the hydraulic network are required to derive a ¾ relationship between surface area and volume at the level of both individual internodes and whole tree structures. If bulk tissue density is constant across branches locally, and metabolic rate is proportional to leaf area, which is in turn proportional to stem surface area, a ¾ relationship between metabolism and mass emerges.

**Self loading**

Next, we consider the addition of biomechanical theory to meet the demands of self loading. Theory linking tree height to stem diameter has long been established based on Euler's buckling model (McMahon & Kronauer 1976; Niklas 1994), which predicts that the maximum height ($l_{max}$) to which an idealized column can be extended scales with its radius ($r$) as

$$l_{max} = c\left(\frac{E}{\rho g}\right)^{1/3} 2r^{2/3} \qquad (5),$$

where $E$ is the modulus of elasticity, $\rho$ is bulk tissue density, $g$ is the acceleration due to gravity, and $c$ is a proportionality constant. $E$, $g$ and $\rho$ are frequently assumed to be constant (Niklas 1994) leading to $l_{max} \propto r^{2/3}$. Many studies have evaluated elastic similarity in large trees and found empirical support, particularly in the larger branches (Holbrook & Putz 1989; West, Brown & Enquist 1999b). However, non-linearity is also a common feature of empirical data. For example, plots of plant height vs. stem diameter are frequently concave on logarithmic axes with slopes typically steeper than the predicted 2/3 at small size scales (Bertram 1989; Niklas 1995; Muller-Landau *et al.* 2006; Enquist *et al.* 2007).

We propose that different parts of the tree branching system may conform to different physical constraints, with small plants, or the peripheral branches of large trees where biomechanical demands are minimal, more consistent with flow similarity ($l \propto r^2$), and the basal branches of large trees more consistent with elastic similarity ($l \propto r^{2/3}$). In a symmetric bifurcating tree, the increase in branch numbers is proportional to $2^n$ where $n$ is the number of branching generations. Peripheral branches thus are expected to exhibit a strong influence on the allometric slopes due to their relative abundance. The exact contribution of each branching generation to the overall scaling exponent depends on tree size and the nature of the $l \propto r^2$ to $l \propto r^{2/3}$ transition (linear or non-linear), the number of branching generations, and the degree of side branching. These factors will differ between species and will ultimately require detailed simulations and/or empirical measurements to determine. Allowing the length-radius scaling to vary within a tree, between trees of differing size, or between species predicts non-linearity in allometric relationships. Table 1 lists predicted curvatures (convex or concave) for each of the six relationships examined here. Representing the length vs. radius scaling as $l \propto r^a$, one can predict a continuum of variability for each allometric relationship (Table 1), and covariation functions for each pairwise combination of exponents (Table S1), all of which are the function of a single parameter (α). Linear regression fits to curved data will often have slopes that fall between the flow similarity and elastic similarity expectations, however, the expected value of the slope will depend strongly on the size range, side branching, and how evenly each size class is sampled.

**Asymmetric branching**

The above theory predicts the dimensions of idealized symmetric branching networks. Real networks however, are usually not so orderly, and branching is commonly asymmetric. Insight can be gained by considering the probability of a branching event. For fractal networks, that probability is scale-free and results in a power law distribution of both lengths and radii. In contrast, biological networks typically involve the acquisition or distribution of resources that occur over finite, and possibly characteristic, length scales. For example, it has recently been shown that leaf vein networks have frequency distributions of vein radii that are well approximated by a power law, and distributions of vein lengths that are better fit by an exponential distribution, suggestive of a characteristic scale (Price, Wing & Weitz 2011). Hence, one can model branching events as a stochastic process with probability $P(l) = pe^{-pl_c}$, with average length or characteristic scale $l_c$ equal to $1/p$.

Modelling branch lengths as a stochastic process raises the question of how empirical data might follow the predictions for a fractal network if they also branch stochastically i.e. asymmetrically. In contrast to previous approaches, the model herein does not require volume to be conserved globally across all branching generations (i.e. $\sum l_k r_k^2$ need not equal $\sum l_{k+1} r_{k+1}^2$). Rather we derive predicted relationships between the basal radius of a subtree and its length, surface area and volume, which should hold for all subtrees within a branching structure regardless of symmetry. Therefore, by examining sub-trees within a network (Bentley *et al.* 2013), i.e. treating each as an individual tree, one can test whether whole trees conform to the expectations developed for ideal symmetric networks as we illustrate below.

**Materials and Methods**

The model described herein makes predictions for geometric, hydraulic and biomechanical scaling relationships in plants and thus numerous tests to evaluate its predictions could be envisioned. We focus here on the network geometry as a first test for the simple reason that if the geometric predictions are not met, subsequent tests of hydraulic or biomechanical predictions are less relevant. To do so we evaluate the allometry of network dimensions in the branches of whole tree saplings, terminal stems, petioles and leaf veins (described below). All individual plant stems, petioles, or veins were approximated as cylinders based on their length and diameter, with surface area and volume for each approximated using standard geometric formulas. The predictions from the above theory were then evaluated in four ways: **1)** by examining standardized major axis (SMA) regression slopes fit to bivariate relationships between length, diameter, surface area and volume and comparing slopes to flow and elastic similarity model predictions; **2)** by examining the frequency distributions of branch lengths and diameters to determine if they are better fit by an exponential or power law model (applicable to two of the datasets, "*plant data*" and "*leaf vein data*", described below); **3)** by examining the daughter/parent branch area ratios (applicable to two of the datasets, "*plant data*" and "*leaf vein data*"), and; **4)** examining the curvature in length-mass-diameter relationships in a large plant allometric dataset. To compare like with like, surface area under elastic similarity was evaluated as the surface area of a cylinder following elastic similarity ($l \propto r^{2/3}$), not proportional to the number of terminal branches as in WBE.

The five datasets used to evaluate the model are described below. In the interest of clarity we will refer to these as "*tree data*", "*stem data*", "*petiole data*", "*leaf vein data*", and "*allometric data*" throughout.

*Tree data:* The length, diameter and connectivity of all stems greater than 1 mm were measured in 19 individual saplings, from four species all within the *Eucalyptus* genus, in the family *Myrtaceae*. The species, with number of individuals in parentheses, were *E. gomphocephala* (6), *E. caesia* (5), *E. diversicolor* (4), and *E. incrassata* (4). These individuals were grown from seed for two years under light shade on the University of Western Australia campus prior to harvest. Individuals were selected to span as wide a range of intraspecific size as possible, with the number of branches per individual ranging from 3 to 59 with a mean of ~31. These data were then examined as bivariate relationships between the individual stem segment dimensions at the individual level, at the species level, and across all species. Data were analysed as "raw data", i.e. individual branch segments, and also within all possible "subtrees", where a subtree is defined as the diameter of a given branch segment, and the total length, total surface area, and total volume of all branch segments distal to that branch segment.

*Stem data:* Terminal stems, defined for this dataset as all stem segments distal to the bud scar from the previous year, were collected from 122 species from the *Banksia* genus (referred to as "stems" throughout) in August 2012. Most species were represented by a single stem, but several species had multiple stems, with a maximum of 44 stems (*B. hewardiana*). Stems with minimal damage or evidence of herbivory and growing in full sunlight were selected. All *Banksia* stems were collected from the Banksia Farm (www.banksiafarm.com.au), a private arboretum containing almost all known members of the *Banksia* genus, and are thus effectively from a common garden. The Banksia Farm is located in Mount Barker (34°37′48″S, 117°40′1″E), situated approximately 370 km south of Perth, Western Australia.

Mount Barker is characterised by a temperate climate, with an annual average high temperature of 20.1°C, an average low of 9.4°C, and a mean annual precipitation of 725 mm.

*Petiole data:* 935 individual leaves from 43 temperate angiosperm species were collected as part of a leaf allometric study in 2007-2008 (full description in; Price *et al.* 2009) for which the petiole dimension data was not analysed or published. For each species, between 18 and 40 (mean = 21.7) individual leaves of increasing size were collected and the length and diameter of their petioles recorded.

*Leaf vein data:* All veins within single leaves from three species; *Banksia victoriae, Hardenbergia comptoniana,* and *Lespedeza cuneata*, were measured as part of a study on the effects of measurement scale on leaf vein dimensions (Price, Munro & Weitz 2014). Leaf subsections were photographed at 5x magnification to reveal minor veins and overlapping images stitched together to form a mosaic image of each entire leaf. The dimensions of all veins in the mosaic leaf images were then measured using the LEAF GUI software (for a full description see; Price, Munro & Weitz 2014).

*Allometric data*: The Sonoran Desert allometric dataset is comprised of plant height, basal stem diameter, and aboveground dry mass for 1509 individuals from 63 species all found growing in the Sonoran Desert region of the southwestern U.S. These data were previously analysed to evaluate covariation in intraspecific allometric relationships, but the interspecific relationships described herein were not published (for a full description see; Price, Enquist & Savage 2007).

*Statistics*

All bivariate relationships were log-transformed prior to analyses. We used standardized major axis regression (SMA) in the software package SMATR to estimate the slopes for all relationships as is common practice in allometric analyses (Warton *et al.* 2006). Frequency distributions of segment lengths and diameters for the *tree data* and *leaf vein data*, were fit with exponential and power-law models following Price et al. (2011). To compare the exponential and power law model fits, we used the method of maximum likelihood to estimate the model parameters and likelihood. We then used a sample size corrected Akaike's information criterion to compare models, $AICc = AIC+(2k(k+1))/(n-k-1)$ (Burnham & Anderson 2002), where $k$ is the number of model parameters, which is 1 for the exponential model, and 2 for the power law model, and $L$ is the likelihood. Table 2 reports corrected AIC scores for the exponential (AICc_E) and power law models (AICc_P) and their relative likelihood, which is the probability that the model with the lower AICc score minimizes information loss (Burnham & Anderson 2002).

**Results**

Standardized major axis regression results for the data presented in Fig. 1 are in Table 1. Slopes are closer to flow similarity than elastic similarity (WBE) predictions in all cases. The confidence intervals for many of the relationships are quite narrow due to the high $R^2$ values and don't always include the predictions of the flow similarity model, but none include the elastic similarity predictions. Note the substantial increase in $R^2$ values when comparing slopes fit to the raw *tree data* and those fit to *subtree data* (rows 8 and 11 respectively in Table 1). In the interest of figure clarity, the vein data for a single leaf only (*B. victoriae*) is plotted in Fig. 1. Bivariate relationships for all three leaves with regression lines and slopes are shown in Fig. S20 with corresponding regression statistics in Table S5. All three leaves had similar slope values for each of the six relationships examined.

Fig. 2a demonstrates how individual branch segments' lengths (*tree data*) and their corresponding basal diameters are poorly correlated (mean $R^2$ of 0.198) with slopes that are both positive and negative. However, if instead of segment length, one examines the total length of all branches distal to a given branch segment (Fig. 2b), and the diameter of that branch segment, the slopes for the relationships tighten considerably around a mean value of 2.05 (mean $R^2$ of 0.875). Figures S1-19 show all six pairwise relationships for each individual tree, both as raw data and as subtrees. The mean $R^2$ for the raw data across all pairwise regressions was 0.48 while the mean for the subtrees was 0.94 (Table S2). Additional statistics for individual and species level regressions for the tree data are presented in Supplementary Tables S2 and S3, respectively.

Table 2 shows that for the *tree data* and *leaf vein data*, frequency distributions of lengths and diameters are always better fit by exponential and power law models, respectively. Fig. 3 shows the frequency distributions of area ratios for the *tree data* and *leaf vein data*, both of which are well approximated by a normal curve and strongly overlap the expectation for area preserving branching. Fig. 4 shows that for allometric relationships between plant height, basal stem diameter and above-ground plant mass, polynomial fits to data display curvature consistent with that predicted in Table 1. SMA regression statistics for the three relationships are in Table S4.

**Discussion**

Collectively, the data presented herein demonstrate that: *i*) scaling relationships consistent with $l \propto r^2$ and $SA \propto V^{3/4}$ are found throughout above-ground plant branching networks; *ii*) area preserving branching is common, consistent with earlier reports (Horn 1971; Bentley *et al.* 2013); *iii*) the frequency distributions of branch lengths are consistent with the expectations of a Poisson process; and *iv*) by examining subtrees within a tree, one can determine if branching is consistent with expectations developed for theoretical symmetric fractal networks. Taken together, the theory and data presented provide strong support for a ¾ scaling of surface area to volume in plant architecture as a proximate driver of metabolic scaling patterns across plants, and we propose an ultimate mechanism, flow similarity, that differs from earlier modelling efforts (West, Brown & Enquist 1999b).

If basal parts of the branching pathway in plants or leaves serve a greater biomechanical role, departure from exact flow similarity model predictions is expected given that allometric relationships will exhibit some non-linearity. While it is possible to predict the direction of the curvature (convex or concave), predicting the exact function, or the slopes of linear models fit to curvilinear data, is challenging, as it will depend on plant size, branch size and attendant biomechanical demands, and the degree of side branching. This may explain why observed slopes are slightly higher than predicted values for some of the relationships in Table1 (columns 4-7). For example, the slopes of lines fit to data in the form of convex curves that have positive first derivatives everywhere, will usually fall between the minimum and maximum derivative values, i.e. between flow similarity and elastic similarity predictions.

The theoretical approach described herein has several advantages over its predecessors. It is a single parameter model that is consistent with known mechanisms of leaf display (along branches, not just at tips), and of fluid loss from xylem (transmural flow) (Zwieniecki *et al.* 2002). The model also operates both at the level of branch internodes and across subtrees within a tree, thus it is relatively easy to evaluate empirically (Supplementary Note 2). Incorporating both flow similarity and elastic similarity into a common framework helps to explain the curvature common to many allometric datasets (Bertram 1989; Niklas 1995; Muller-Landau *et al.* 2006; Enquist *et al.* 2007), why linear fits to curvilinear data fall between the different model predictions, and why interspecific allometric relationships covary (Price, Enquist & Savage 2007). However, the specific covariation functions described here (Table S1) differ from those predicted in earlier work (Price, Enquist & Savage 2007). This is because the predicted branch dimensions differ between the models, and because under flow similarity, surface area is proportional to branch surface area whereas under the Price et al. (2007) approach (based on the WBE framework), surface area is proportional to the number of terminal branches.

The model developed here is an idealized abstraction, much like its conceptual predecessors (Shinozaki *et al.* 1964; West, Brown & Enquist 1997; West, Brown & Enquist 1999b; Savage *et al.* 2010) and its simplifying assumptions will generally be more valid at the species or genus level, rather than at the family level or higher. Empirical validation of the model will depend strongly on the trait in question (i.e. bulk density, leaf size or stem specific conductivity), the amount of trait variance for the clade in question, and whether such variance changes systematically with plant size (Price *et al.* 2014). The theory assumes that leaf area is proportional to stem surface area in terminal branches, but in many species, leaves are ephemeral and thus total leaf area produced over a growing season may be more tightly correlated with stem surface area rather than the leaf area or number of leaves found at any one time.

Published data for the logarithmic relationship between above ground dark respiration and plant mass suggest that the scaling is isometric at small sizes (Reich *et al.* 2006) shifting to a slope close to ¾ at large sizes (Mori *et al.* 2010). This may result from the fact that in small seedlings and saplings, most or all tissue is metabolically active, and total respiration is not limited by the stem surface area available for leaf display. However, as trees grow larger, an increasing proportion of a trees' total mass is composed of tissue that has low or no metabolic activity, and respiration will increasingly be dominated by leaves, and to a lesser extent, the sapwood, the active cambium layer and the phloem, all of which are expected to be approximately proportional to branch surface area, causing empirical slopes to shift towards the theoretical relationships described herein.

Velocity preservation and the conservation of volume flow rate are intuitive and arguably parsimonious model constraints. However, more integrative traits that reflect network efficiency may be the targets upon which natural selection is ultimately acting. For example, for a given $Q$ and $\Delta P$, and a constant sapwood area fraction, a branching that follows $l \propto r^2$ would conserve sapwood specific conductivity ($K_S$), defined as $K_S = \frac{Ql}{\Delta P r_s^2}$, where $r_s$ is the sapwood radius. While space precludes a detailed exploration of variability in stem conductivity within and across tree branches, there is evidence to suggest that in the absence of environmentally driven variation, or branch order/path length dependent effects, sapwood specific conductivity may be a conserved species-specific trait (McDowell *et al.* 2002; Sellin, Rohejarv & Rahi 2008). Specific conductivity needn't be conserved throughout the entire plant branching network for flow similarity to have a strong influence on the allometry of metabolism, again due to the numerical dominance of the terminal parts of the branching pathway.

Area preserving branching has strong empirical support among external tree branches (Horn 2000; Bentley *et al.* 2013), and across vessel bundles in leaves (Price, Knox & Brodribb 2013), but the physical processes underlying this principle are not yet fully established. Published data suggest that velocity can increase, decrease or not vary statistically as a function of branch diameter within and across species (McCulloh & Sperry 2005; Savage *et al.* 2010). Thus the extent to which velocity preservation can be invoked as a global constraint remains an open question. It is easy to envision that sap velocity may be more constrained locally as rapid changes in velocity over short distances would seem disadvantageous. Indeed velocity measures for similar sized branches within the same species are often quite close to one another (Savage *et al.* 2010). In a recent summary of existing data, Savage et al. (2010) found no significant relationship between branch diameter and maximum sap velocity in 8 out of 12 species. In light of the relatively low number of species for which these variables have been measured, the relationships between branch geometry, topology, and fluid dynamics warrant further inquiry.

We have invoked a linear relationship between internal and external branching locally as a simplifying model assumption. However, tapering of conduit dimensions is well supported empirically at the scale of entire trees (Ewers & Zimmerman 1984; Anfodillo *et al.* 2006; Coomes, Jenkins & Cole 2007; Sellin, Rohejarv & Rahi 2008; Savage *et al.* 2010), and this would imply a non-linear relationship between internal and external branching (Savage et al. 2010). It can be shown that our model, with its assumptions of area-preserving branching and flow similarity, and its data-supported prediction of volume to surface-area scaling are consistent with conduit dimensions tapering if the pressure drop across branching generations varies in a certain way from parent to daughter branches (Supplementary Note 1).

We highlight the distinction between the proximate and ultimate mechanisms we describe for the quarter power scaling of plant metabolism. The proximate mechanism is the $3/4^{th}$ scaling of branch surface area to volume which appears to have strong empirical support in our data. Subsequent characterization of this pattern in other clades and larger trees is required to understand the full scope of this mechanism. We offer the maintenance of flow similarity as one possible ultimate mechanism, particularly if the terminal branches in plants have a strong influence on metabolic scaling exponents, but recognize that empirical support are needed for the simplifying assumptions of velocity preservation and a constant pressure drop, or variable pressure drop with conduit tapering (Supplemental Note 1). Further sensitivity analyses exploring the relaxation of these assumptions will help to determine how they potentially influence metabolic scaling patterns.

If flow similarity, as reflected in $l \propto r^2$ scaling, underlies or contributes to quarter power scaling in plants, it is natural to speculate as to why it has remained hidden. This may be due to several factors. First, because of the stochastic nature of length branching, raw plots of branch length vs. diameter measures will exhibit poor correlations with highly variable slopes, effectively obscuring the $l \propto r^2$ signal that may be revealed by examining subtrees (Figs. 2, S1-S19). Second, a focus on the scaling of tree height with stem diameter in large trees (McMahon & Kronauer 1976; West, Brown & Enquist 1999b) may have driven a search for explanations that exhibit only $l \propto r^{2/3}$ scaling, which may apply to large branches or tree height in large trees, but does not explain the $l \propto r^2$ scaling in abundant branch ends, or the non-linearity common to many datasets: total plant height and total branch path length are both qualitatively and quantitatively different phenomena. Lastly, most theoretical attempts have searched for global optima, not allowing for the fact that different parts of networks may be optimized to perform different functions (Price, Knox & Brodribb 2013).

Some have questioned whether allometric patterns in biology such as "Kleiber's law" are even laws at all, noting the variability in both intraspecific and interspecific scaling exponents observed in empirical data (Dodds, Rothman & Weitz 2001; Glazier 2006). Large collections of interspecific data, and several meta-analyses of intraspecific data have shown that empirical data do often (but not always) have slopes that cluster around values close to the canonical ¾ (Niklas & Enquist 2001; Brown *et al.* 2004; Savage *et al.* 2004; Mori *et al.* 2010) albeit with curvature in some datasets (Savage *et al.* 2004; Mori *et al.* 2010). Questions of whether or not collections of allometric data showing scaling exponents near ¼ constitute a "biological law" are largely semantic in nature and not easily answered to the satisfaction of all. Perhaps a more productive approach would be to ask whether scaling exponents that are some multiple of ¼ are common enough that they might emerge from a common mechanism such as that described herein.

**Acknowledgements**

CAP wishes to acknowledge the support of an Australian Research Council (ARC) Discovery Early Career Researcher Award (DECRA) and a fellowship from the National Institute for Mathematical and Biological Synthesis (NIMBioS, USA). Jane Price, Susan Miller and Cathy Angel provided helpful comments on an earlier version of this manuscript. Thanks to Christina Wilson, Farhad Amani, Sarah-Jane Knox, and Yue Wang for data collection, and to Jason Hamer for raising the eucalypt saplings. Thanks also to the Banksia Farm for allowing us to collect branch samples.

| Y-variable | Length | Surface Area | Diameter | Length | Diameter | Length |
|---|---|---|---|---|---|---|
| X-variable | Diameter | Volume | Volume | Volume | Surface Area | Surface Area |
| Expression | $L=D^{\alpha}$ | $SA=V^{(\alpha+1)/(\alpha+2)}$ | $D=V^{1/(\alpha+2)}$ | $L=V^{\alpha/(\alpha+2)}$ | $D=SA^{1/(\alpha+1)}$ | $L=SA^{\alpha/(\alpha+1)}$ |
| Flow Similarity | 2 | 3/4 | 1/4 | 1/2 | 1/3 | 2/3 |
| Elastic Similarity | 2/3 | 5/8 | 3/8 | 1/4 | 3/5 | 2/5 |
| Changing Exponent | 2 to 2/3 | 3/4 to 5/8 | 1/4 to 3/8 | 1/2 to 1/4 | 1/3 to 3/5 | 2/3 to 2/5 |
| Curvature | Concave | Concave | Convex | Concave | Convex | Concave |
| *Tree data* slope (raw data) (588) | -1.639 | 0.749 | 0.389 | 0.638 | 0.524 | 0.858 |
| *Tree data* slope CI's | -1.777 to -1.511 | 0.723 to 0.763 | 0.370 to 0.410 | 0.599 to 0.679 | 0.489 to 0.562 | 0.823 to 0.896 |
| *Tree data* $R^2$ | 0 | 0.888 | 0.593 | 0.395 | 0.263 | 0.726 |
| *Tree data* slope (subtrees) (588) | 2.06 | 0.798 | 0.314 | 0.648 | 0.394 | 0.813 |
| *Tree data* slope CI's | 1.998 to 2.131 | 0.791 to 0.805 | 0.308 to 0.321 | 0.635 to 0.662 | 0.384 to 0.404 | 0.802 to 0.823 |
| *Tree data* $R^2$ | 0.842 | 0.989 | 0.928 | 0.932 | 0.901 | 0.973 |
| *Stem data* slopes (436) | 1.978 | 0.763 | 0.288 | 0.567 | 0.376 | 0.743 |
| *Stem data* slope CI's | 1.84 to 2.126 | 0.751 to 0.776 | 0.278 to 0.298 | 0.542 to 0.593 | 0.358 to 0.396 | 0.720 to 0.766 |
| *Stem data* $R^2$ | 0.506 | 0.972 | 0.879 | 0.769 | 0.768 | 0.892 |
| *Petiole data* slopes (955) | 1.979 | 0.759 | 0.282 | 0.558 | 0.372 | 0.735 |
| *Petiole data* slope CI's | 1.88 to 2.084 | 0.751 to 0.767 | 0.274 to 0.29 | 0.542 to 0.574 | 0.358 to 0.386 | 0.721 to 0.749 |
| *Petiole data* $R^2$ | 0.348 | 0.972 | 0.797 | 0.793 | 0.648 | 0.91 |
| *Leaf vein data* slopes (9502) | 1.939 | 0.7765 | 0.3418 | 0.6628 | 0.4402 | 0.8537 |
| *Leaf vein data* slope CI's | 1.901 to 1.979 | 0.772 to 0.781 | 0.337 to 0.346 | 0.654 to 0.672 | 0.433 to 0.448 | 0.846 to 0.862 |
| *Leaf vein data* $R^2$ | 0.011 | 0.916 | 0.565 | 0.538 | 0.279 | 0.808 |

**Table 1.** Predicted and observed relationships between length, diameter, surface area and volume. Y and X variables are listed in the top two rows. An expression for each relationship is in the third row, where α is the length to diameter exponent which is equal to 2 under flow similarity. Rows 4-7 represent the predictions for flow similarity, elastic similarity, the change in exponent expected going from small to large plants (i.e. from flow to elastic similarity), and the expected curvature from such a relationship. Rows 8-22 represent the observed slopes, 95% slope CI's and $R^2$ for each relationship, for the *tree data*, *tree data* subtrees, *stem data*, *petiole data* and *leaf vein data* The observed slope values are shaded to facilitate comparison between groups and predictions.

| Organ | Species | Dimension | Exponential | Power Law | AICc_P | AICc_E | Relative Likelihood | Sample Size |
|---|---|---|---|---|---|---|---|---|
| leaf veins | B. victoriae | lengths | Y | | -7.97E+03 | -1.20E+04 | 0 | 9502 |
| leaf veins | B. victoriae | diameters | | Y | -5.30E+04 | -3.64E+04 | 0 | 9502 |
| leaf veins | H. comptoniana | lengths | Y | | -2.11E+04 | -3.18E+04 | 0 | 24148 |
| leaf veins | H. comptoniana | diameters | | Y | -1.41E+05 | -1.10E+05 | 0 | 24148 |
| leaf veins | L. cuneata | lengths | Y | | -3.95E+03 | -6.58E+03 | 0 | 3096 |
| leaf veins | L. cuneata | diameters | | Y | -2.64E+04 | -2.08E+04 | 0 | 3096 |
| tree branches | E. gomphocephela | lengths | Y | | 1.61E+03 | 1.33E+03 | 1.33E-61 | 203 |
| tree branches | E. gomphocephela | diameters | | Y | -91.78 | -88.09 | 0.16 | 203 |
| tree branches | E. ceasia | lengths | Y | | 629.75 | 541.61 | 7.25E-20 | 94 |
| tree branches | E. ceasia | diameters | | Y | -105.09 | -97.55 | 0.02 | 94 |
| tree branches | E. diversicolor | lengths | Y | | 1.66E+03 | 1.38E+03 | 2.13E-60 | 203 |
| tree branches | E. diversicolor | diameters | | Y | -111.22 | -108.38 | 0.24 | 203 |
| tree branches | E. incrassata | lengths | Y | | 663.87 | 581.04 | 1.03E-18 | 88 |
| tree branches | E. incrassata | diameters | | Y | -123.24 | -78.83 | 2.27E-10 | 88 |

**Table 2.** For the two datasets that contained full hierarchical trees, *leaf vein data* and *tree data*, each species was tested to determine if the distribution of vein or branch segment lengths and diameters were better fit by a exponential model (column 4) or a power law model (column 5). In all cases, length distributions were better fit by an exponential model, and diameter distributions better fit by a power law (indicated by the letter "Y"). Columns 6 and 7 represent the size corrected AIC score for the power law and exponential models respectively, with their relative likelihood in column 8 followed by the sample size.

**Figure Legends**

**Figure 1**. Bivariate plots of the dimensions of the *tree data*, *stem data*, *petiole data* and *leaf vein data*, plotted on a common axis for pairwise relationships between length, diameter, surface area and volume (Panels A-F). SMA regression statistics for all relationships are presented in Table 1. *Tree data* is based on subtrees as described in the methods. While length-diameter relationships are characterized by lower coefficients of determination ($R^2$), surface area-volume relationships are tightly correlated.

**Figure 2**. Allometric relationships between stem segment length and diameter in raw *tree data* (Panel A), and the same relationship in subtrees within each of 19 trees (Panel B, see Methods). The raw data correlations are highly variable with both positive and negative slopes (mean slope = -0.95 and slope standard deviation 1.54), however, subtree regressions converge toward the predicted value of 2 (mean subtree slope = 2.05 and slope standard deviation = 0.24). This demonstrates how by examining subtrees within a tree can help determine if branching conforms to the expectations developed for symmetric fractal trees.

**Figure 3**. Frequency distributions for *tree data* and *leaf vein data* branch area ratios. Each distribution is well approximated by a normal curve and includes the expectation for area-preserving branching (black vertical line). The mean *tree data* branch area ratio (red vertical line) is slightly lower than the expected value, while the *leaf vein data* area ratio is slightly higher.

**Figure 4**. Allometric relationships between plant height, basal stem diameter and above ground plant mass from a compilation of Sonoran Desert plant *allometric data*: The red line in each plot represents a 2$^{nd}$ order polynomial fit to the data to determine the curvature, and the light blue line represents SMA regression fit. Polynomial curvatures are consistent with those predicted (Table 1), concave in panels A and B, and convex in panel C. Regression slopes for panels A (1.1) and B (0.43) fall in between flow and elastic similarity predictions as might be expected, and the slope for Panel C (0.42) is just outside of this range (see Table 1).

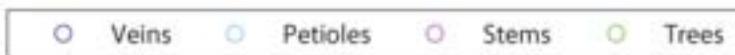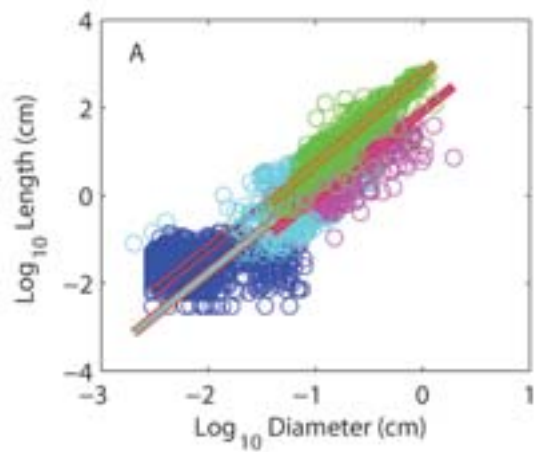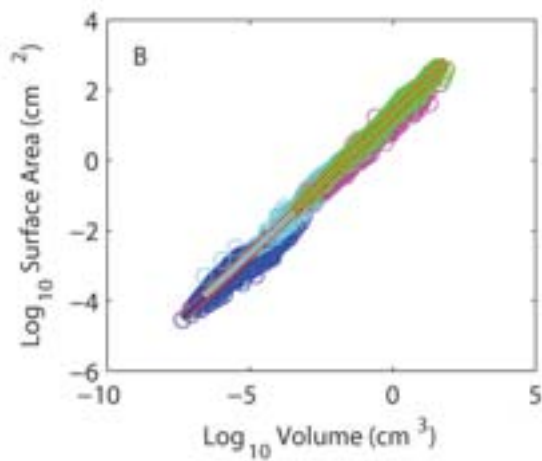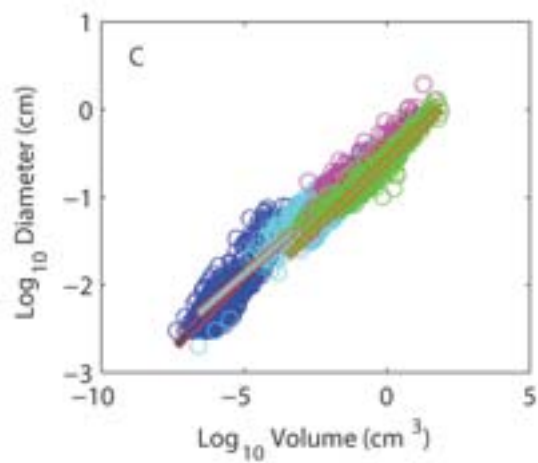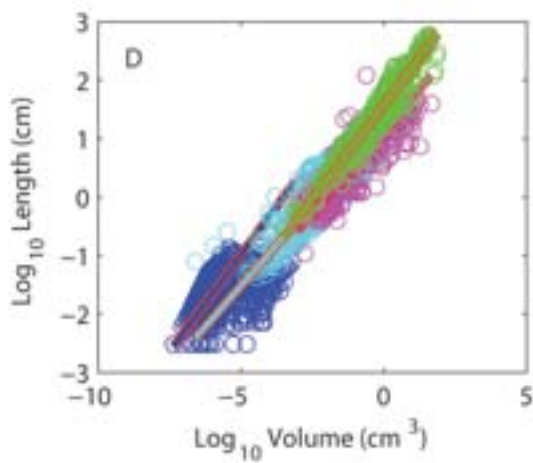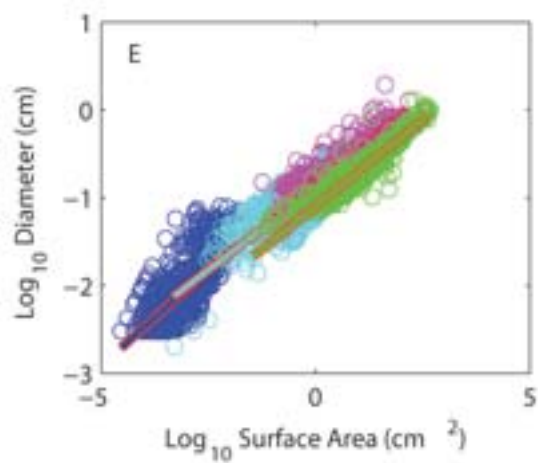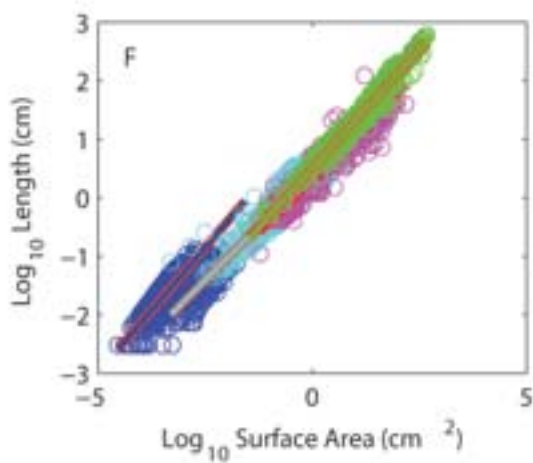

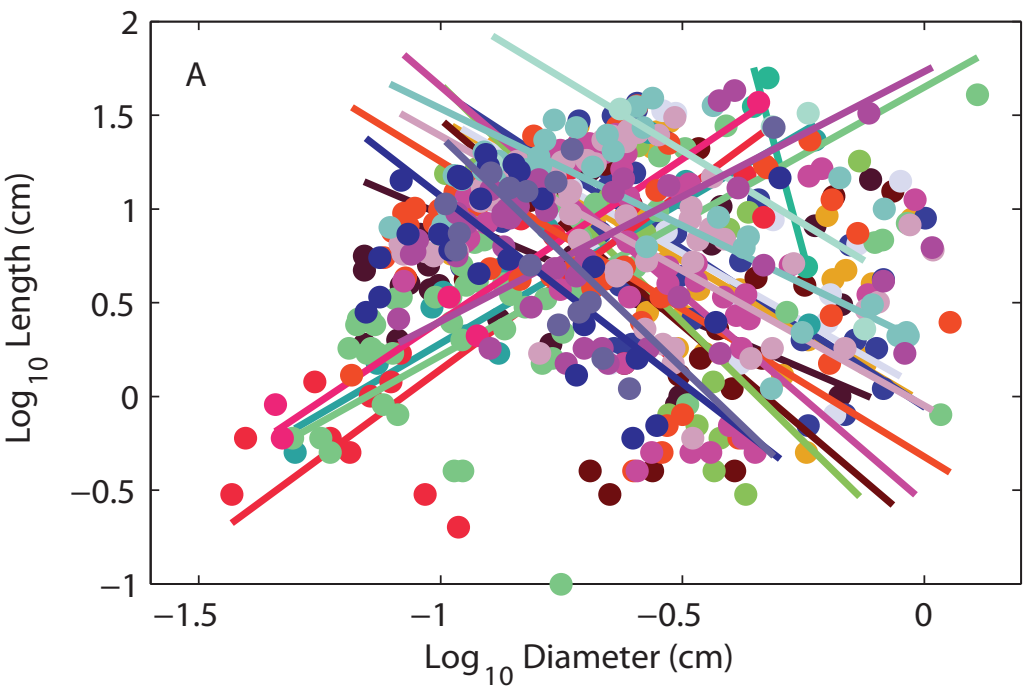

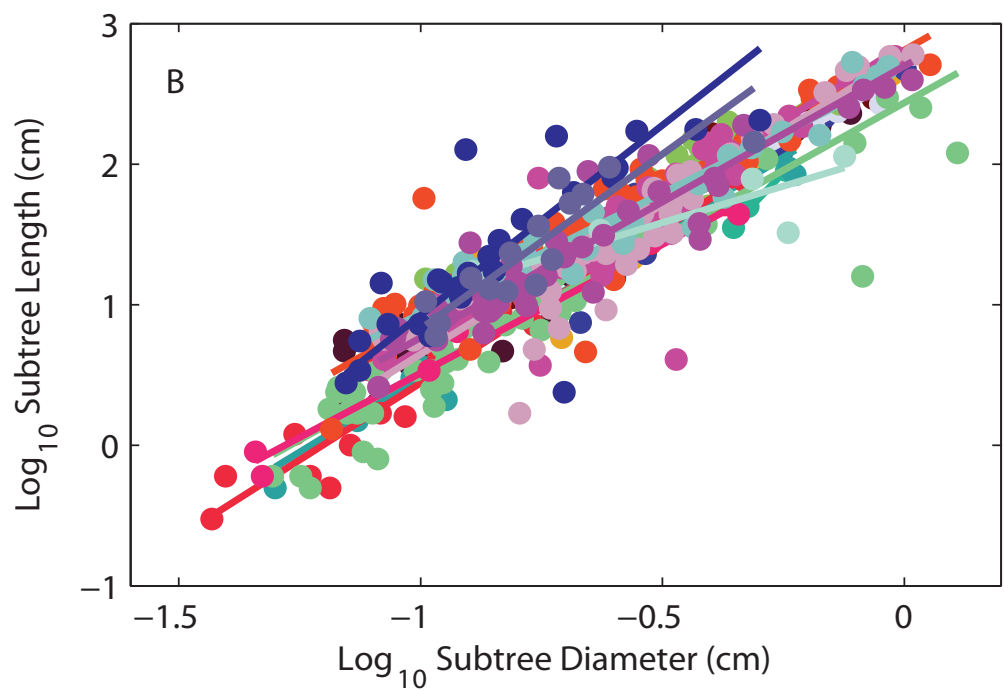

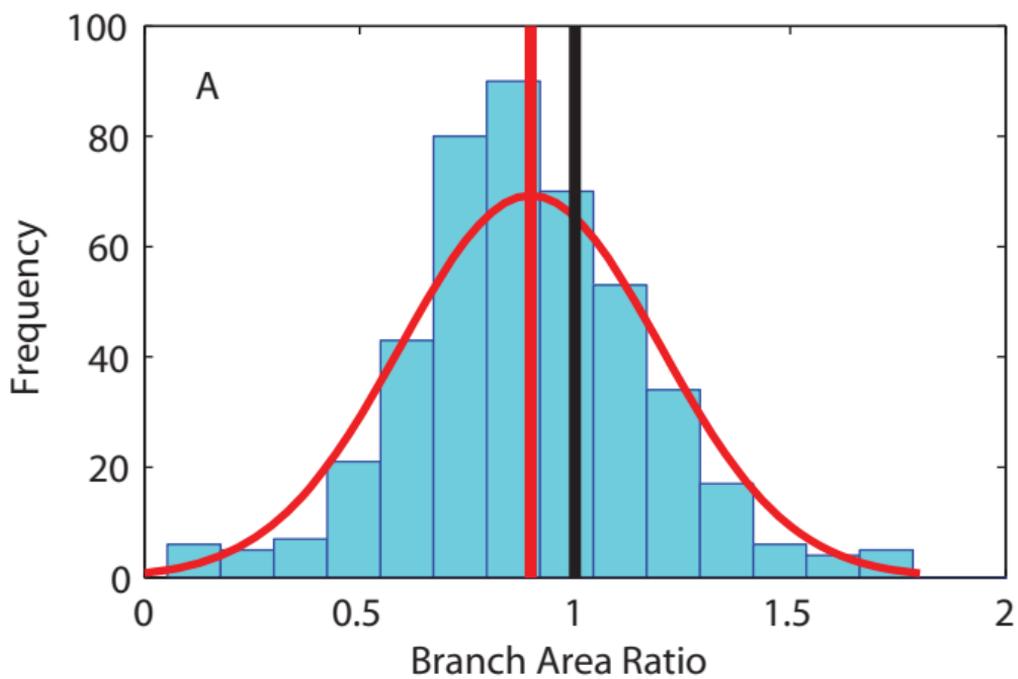
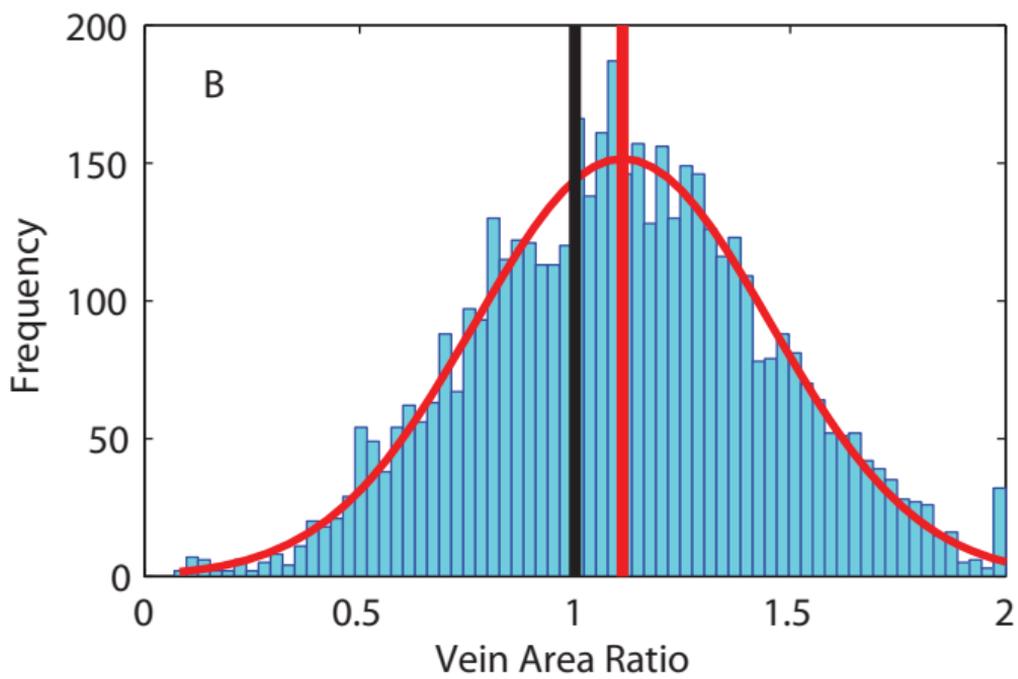

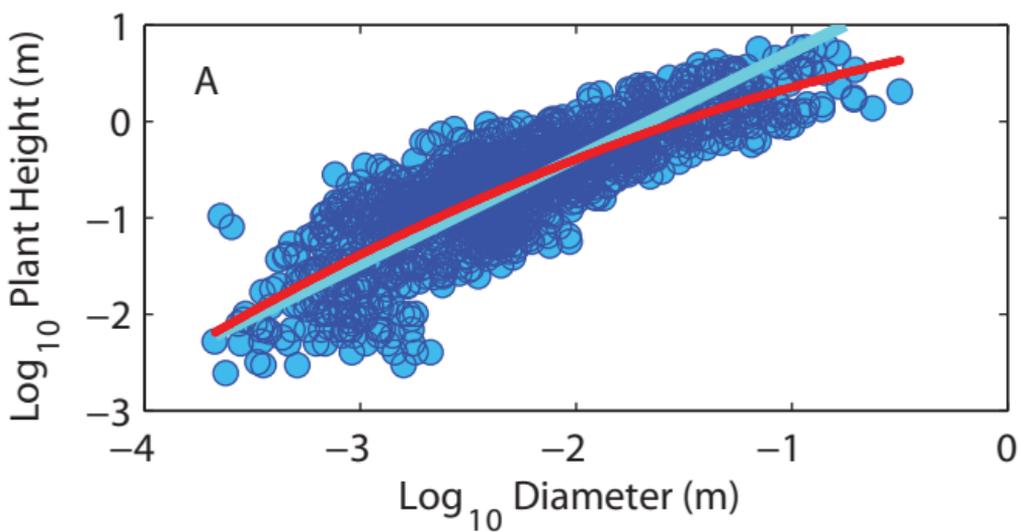
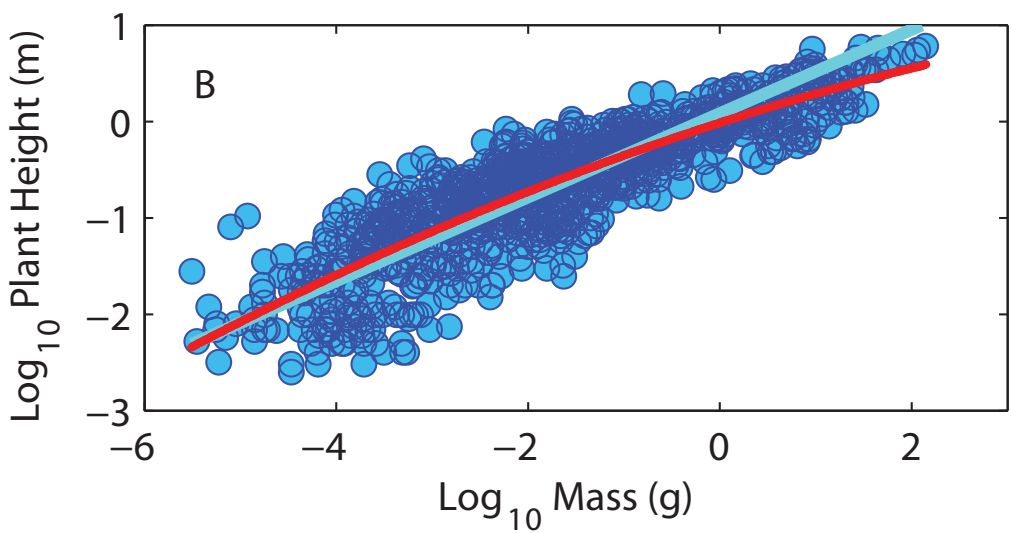
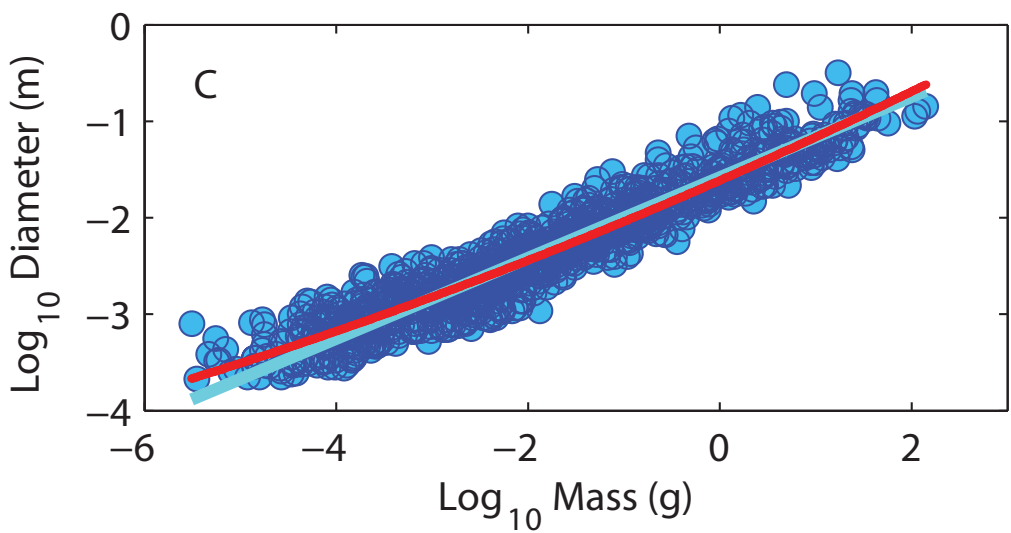

Supplementary Note 1

Our basic model assumes area-preserving branching and flow similarity (conservation of volumetric flow rate with branching), which together imply constant flow velocity through the branching structure. Perhaps more tentatively, we also assume a constant pressure drop and a linear relationship between internal and external branching characteristics (eg. branching is area-preserving both for external branching structure and for internal branching conduits), and show that this leads to a prediction of surface-area scaling with volume to the power ¾. We then show that this prediction to be well-supported by empirical data.

Here we show that our model, with its basic assumptions of area-preserving external branching and flow similarity, is still consistent with both the data-supported prediction of surface-area scaling with volume to the power ¾ and the conduit dimensions tapering modelled by Savage et al. (2010) if pressure drop varies in a certain way from parent to daughter branches.

Savage et al. (2010) show that theoretical optimality arguments lead to

$$r_{int\ k} \propto r_{ext\ k}^{\frac{1}{3}} \qquad (1)$$

and

$$N_{int\ k} \propto r_{ext\ k}^{\frac{4}{3}} \qquad (2)$$

where $r_{int\ k}$ and $r_{ext\ k}$ are the radii of the internal conduits and external branches at the $k^{th}$ branching generation and $N_{int\ k}$ is number of xylem conduits at level $k$.

If we also assume area-preserving bifurcating external branching

$$r_{ext\ k} \propto \sqrt{2}\ r_{ext\ k+1} = 2^{\frac{1}{2}} r_{ext\ k+1} \qquad (3)$$

where $k+1$ is the order of branching more distal than $k$, as per the usual labelling, then it follows that

$$r_{int\ k} = 2^{\frac{1}{6}} r_{int\ k+1} \qquad (4)$$

and

$$N_{int\ k} = 2^{\frac{2}{3}} N_{int\ k+1} \qquad (5)$$

If we then assume the key relationship we obtain from our basic model, which is the key result leading to the other relationships supported by the empirical data

$$l_{int\ k} = 2\ l_{int\ k+1} \qquad (6)$$

and assume, like Savage et al. (2010), that internal lengths mirror external lengths,

$$l_{ext\ k} = 2\ l_{ext\ k+1} \qquad (7)$$

then the total volumetric flow at order $k$

$$TQ_k = N_{int\ k} \frac{\pi r_{int\ k}^4 |\Delta P_k|}{8\eta l_{int\ k}}$$

$$= 2^{\frac{2}{3}} N_{int\ k+1} \frac{\pi r_{int\ k+1}^4 2^{\frac{2}{3}} |\Delta P_k|}{8\eta \cdot 2 l_{int\ k}} \qquad \text{(from 4, 5, 6)}$$

$$= 2^{\frac{1}{3}} N_{int\ k+1} \frac{\pi r_{int\ k+1}^4 |\Delta P_k|}{8\eta \cdot l_{int\ k}} \qquad (8)$$

and if volumetric flow is conserved across generations then

$$TQ_{k+1} = N_{int\ k+1} \frac{\pi r_{int\ k+1}^4 |\Delta P_{k+1}|}{8\eta l_{int\ k+1}} = TQ_k = 2^{\frac{1}{3}} N_{int\ k+1} \frac{\pi r_{int\ k+1}^4 |\Delta P_k|}{8\eta \cdot l_{int\ k}}$$

and thus

$$|\Delta P_{k+1}| = 2^{\frac{1}{3}} |\Delta P_k| \qquad (9)$$

or as a pressure gradient instead of a pressure drop

$$\frac{|\Delta P_{k+1}|}{l_{int\ k+1}} = \frac{2^{\frac{1}{3}} |\Delta P_k|}{2^{-1} l_{int\ k}} = 2^{\frac{4}{3}} \frac{|\Delta P_k|}{l_{int\ k}} \qquad (10)$$

This means that the assumptions of area-preserving external branching and flow similarity can be compatible with both the conduit tapering modelled in Savage et al. 2010 and the length to radius squared and surface area to volume to ¾ scalings that follow from (3) and (7) and find support in our empirical data, if pressure varies within the branching structure according to (9) and (10). Note that is a steeper pressure drop/gradient than the original assumptions without tapering where pressure drop is constant, and thus pressure gradient doubles with each successive branching order as length halves (6).

Supplementary Note 2

Here we show that for a symmetric bifurcating tree structure:

1. length scales with radius squared through the whole structure at the scale of individual internodes
2. surface area scales with volume to the power ¾ through the whole structure at the scale of individual internodes
3. the total surface area of the tree scales with the total volume of the tree to the power ¾ as the tree increases in size
4. the total length of the tree scales with the basal radius of the tree squared as the tree increases in size

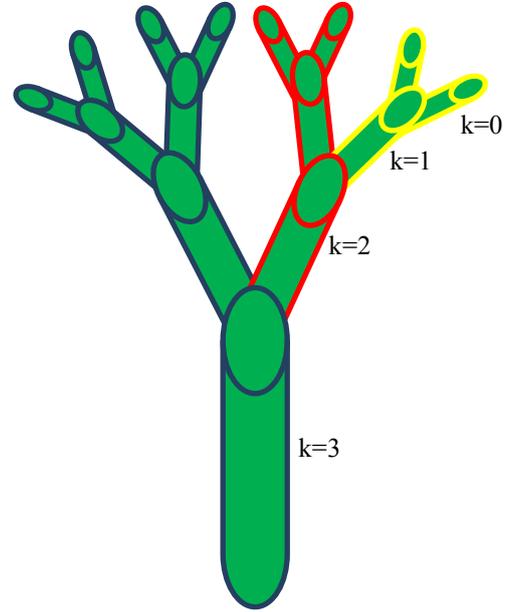

Let $v_k$, $s_k$, $l_k$, and $r_k$ be the volume, surface area, length and radius of an internode at the $k^{th}$ level within a symmetric bifurcating tree structure, where the indexing of orders goes from the tips (order 0) to the base (order N), as shown in the figure ie the opposite direction to the normal labelling used in other parts of this paper. Let $V_k$, $S_k$, $L_k$, and $R_k$ then be the total volume, surface area, length and radius of all internodes at the $k^{th}$ level within the structure, and $TV_k$, $TS_k$, and $TL_k$ be the total volume, surface area and length of a subtree of order k ie the total volume (surface area/length) of a $k^{th}$ internode and all internodes distal to that internode.

If we assume $Q = \frac{\pi r^4 |\Delta P|}{8\eta l}$

then $l = \frac{Cr^4}{Q}$

where $C = \frac{\pi |\Delta P|}{8\eta}$.

If we also assume that branching is area preserving, that is $r_k = \sqrt{2} r_{k-1}$, and viscosity and pressure drop are the same in adjacent internodes as previously discussed, then we can establish the following relationships between branches at the k-1$^{th}$ and k$^{th}$ levels

$$l_{k-1} = \frac{Cr_{k-1}^4}{Q_{k-1}} = \frac{C(\frac{1}{\sqrt{2}} r_k)^4}{\frac{1}{2} Q_k} = \frac{1}{2} \frac{Cr_k^4}{Q_k} = \frac{1}{2} l_k \qquad (1)$$

and therefore

$$\frac{l_k}{r_k^2} = \frac{(\frac{1}{2})^k l_0}{\left(\left(\frac{1}{\sqrt{2}}\right)^k r_0\right)^2} = \frac{l_0}{r_0^2} \tag{2}$$

and so length scales with radius squared through the whole structure at the scale of individual internodes.

Then similarly for internode volumes

$$v_{k-1} = \pi r_{k-1}^2 l_{k-1} = \pi (\frac{1}{\sqrt{2}} r_k)^2 \frac{1}{2} l_k = \frac{1}{4} \pi r_k^2 l_k = \frac{1}{4} v_k \tag{3}$$

and surface areas

$$s_{k-1} = 2\pi r_{k-1} l_{k-1} = 2\pi \frac{1}{\sqrt{2}} r_k \frac{1}{2} l_k = \frac{1}{2\sqrt{2}} s_k = 2^{-\frac{3}{2}} s_k \tag{4}$$

And therefore, for any k,

$$\frac{v_{k-1}^{\frac{3}{4}}}{s_{k-1}} = \frac{\left(\frac{1}{4}v_k\right)^{\frac{3}{4}}}{\left(2^{-\frac{3}{2}}s_k\right)} = \frac{2^{-2\times\frac{3}{4}} v_k^{\frac{3}{4}}}{2^{-\frac{3}{2}} s_k} = \frac{v_k^{\frac{3}{4}}}{s_k} \tag{5}$$

and so surface area scales with volume to the power ¾ through the whole structure at the scale of individual internodes.

From Eqn 3, and because there are twice as many branches at order k-1 compared to order k

$$V_{k-1} = \frac{1}{2}V_k \tag{6}$$

Similarly, Eqn 4 says that,

$$s_{k+1} = 2\sqrt{2}s_k$$

and because there are twice as many branches at order k-1 compared to order k

$$S_{k-1} = \frac{1}{\sqrt{2}} S_k \tag{7}$$

It follows from Eqn 6 that

$$TV_k = V_k + V_{k-1} + V_{k-2} + \cdots + V_1 + V_0$$

$$= V_k + \frac{1}{2}V_k + \left(\frac{1}{2}\right)^2 V_k + \cdots + \left(\frac{1}{2}\right)^{k-1} V_k + \left(\frac{1}{2}\right)^k V_k$$

$$= V_k \left(\frac{1-\left(\frac{1}{2}\right)^k}{1-\frac{1}{2}}\right) = 2V_k \left(1-\left(\frac{1}{2}\right)^k\right) = 2v_k \left(1-\left(\frac{1}{2}\right)^k\right) \tag{8}$$

and from Eqn 7 that

$$TS_k = S_k + S_{k-1} + S_{k-2} + \cdots + S_1 + S_0$$

$$= S_k + \frac{1}{\sqrt{2}}S_k + \left(\frac{1}{\sqrt{2}}\right)^2 S_k + \cdots + \left(\frac{1}{\sqrt{2}}\right)^{k-1} S_k + \left(\frac{1}{\sqrt{2}}\right)^k S_k$$

$$= S_k \left(\frac{1-\left(\frac{1}{\sqrt{2}}\right)^k}{1-\frac{1}{\sqrt{2}}}\right) = s_k \left(\frac{1-\left(\frac{1}{\sqrt{2}}\right)^k}{1-\frac{1}{\sqrt{2}}}\right) \tag{9}$$

We then consider the proportional difference in total volume between a tree of order k+1 and another of order k.

$$\frac{TV_{k+1}}{TV_k} = \frac{2TV_k + 4v_k}{TV_k} = 2 + \frac{4v_k}{2v_k(1-\frac{1}{2}^k)}$$

$$= 2 + \frac{2}{(1-\left(\frac{1}{2}\right)^k)} \tag{10}$$

which tends towards 4 as k becomes large.

We similarly consider the proportional difference in total surface area between a tree of order k+1 and another of order k.

$$\frac{TS_{k+1}}{TS_k} = \frac{2TS_k + 2\sqrt{2}s_k}{TS_k} = 2 + \frac{2\sqrt{2}s_k}{s_k\left[\frac{1-\frac{1}{\sqrt{2}}^k}{1-\frac{1}{\sqrt{2}}}\right]}$$

$$= 2 + \frac{2\sqrt{2}\left(1 - \frac{1}{\sqrt{2}}\right)}{1 - \left(\frac{1}{\sqrt{2}}\right)^k}$$

which tends towards

$$2 + 2\sqrt{2}\left(1 - \frac{1}{\sqrt{2}}\right) = 2\sqrt{2}$$

as k becomes large.

And so as k becomes large

$$\frac{\log(\frac{TV_{k+1}}{TV_k})}{\log(\frac{TS_{k+1}}{TS_k})} \rightarrow \frac{\log 4}{\log 2\sqrt{2}} = \frac{2\log 2}{\frac{3}{2}\log 2} = \frac{4}{3} \tag{11}$$

which means that the total volume of the tree scales with the total surface area of the tree to the power 4/3 as the tree increases in size, or in other words the total surface area of the tree scales with the total volume of the tree to the power ¾ as the tree increases in size.

Furthermore, from Eqn 1, combined with the fact that there are twice as many internodes at level k compared to level k-1

$$L_k = L_k \tag{12}$$

and so

$$TL_k = L_k + L_{k-1} + L_{k-2} + \cdots + L_1 + L_0$$
$$= L_k + L_k + L_k + \cdots + L_k + L_k$$
$$= kL_k = kl_k \tag{13}$$

We can then consider the proportional difference in total length between a tree of order k+1 and another of order k.

$$\frac{TL_{k+1}}{TL_k} = \frac{2TL_k + 2l_k}{TL_k} = 2 + \frac{2l_k}{kl_k} = 2 + \frac{2}{k} \tag{14}$$

which tends towards 2 as k becomes large.

And so as k becomes large, the scaling exponent between the total length of a tree and its basal radius

$$\frac{\log\left(\frac{TL_{k+1}}{TL_k}\right)}{\log\left(\frac{r_{k+1}}{r_k}\right)} \to \frac{\log(2)}{\log(\sqrt{2})} = 2 \tag{15}$$

which means that the total length of the tree scales with the basal radius of the tree squared as the tree increases in size.

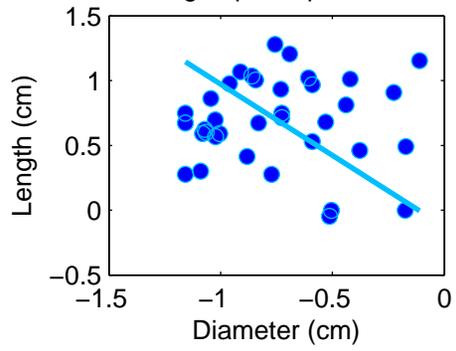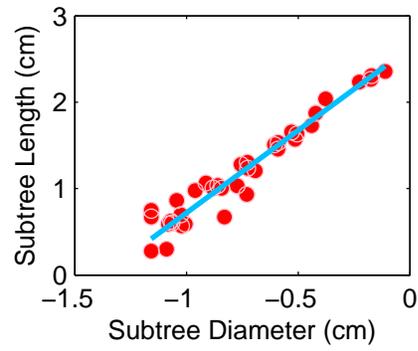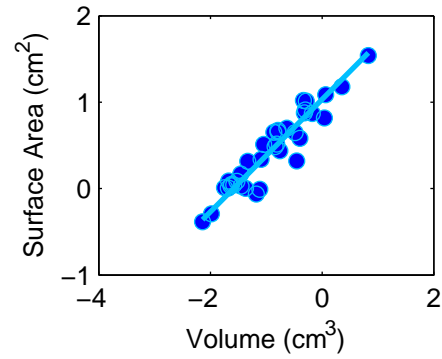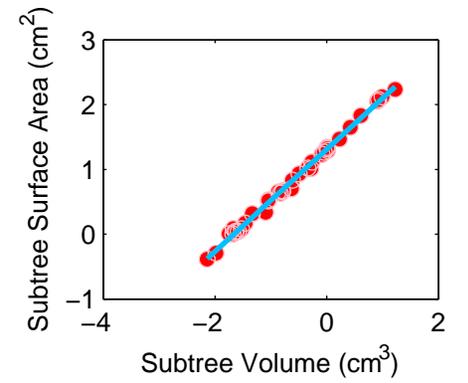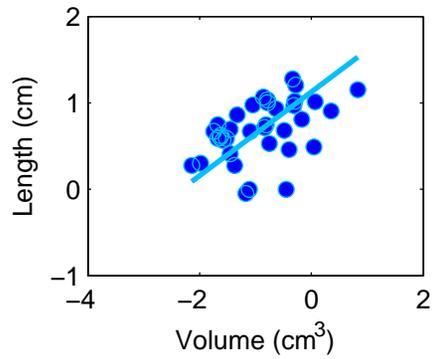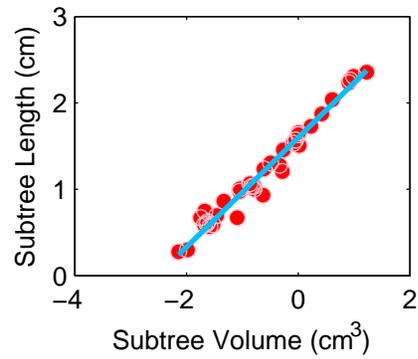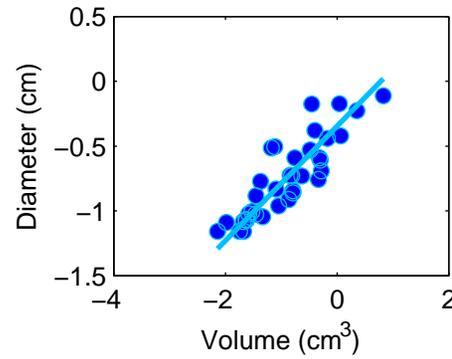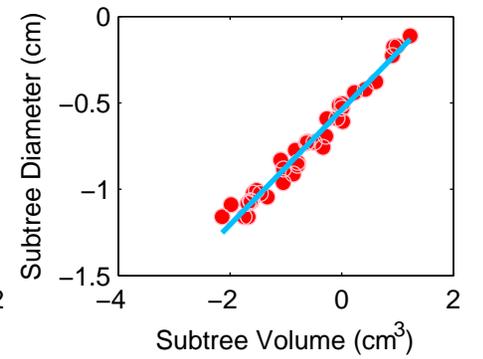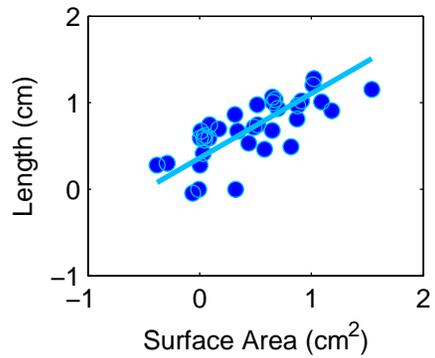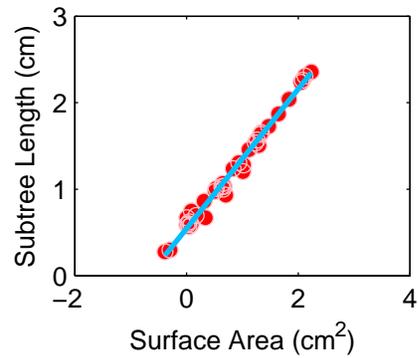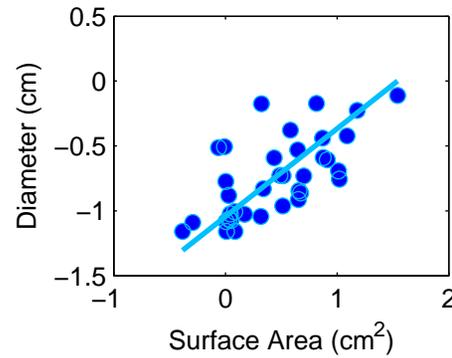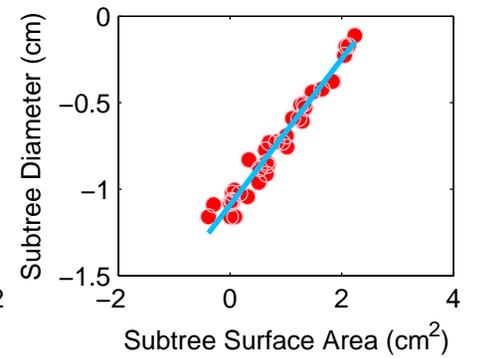

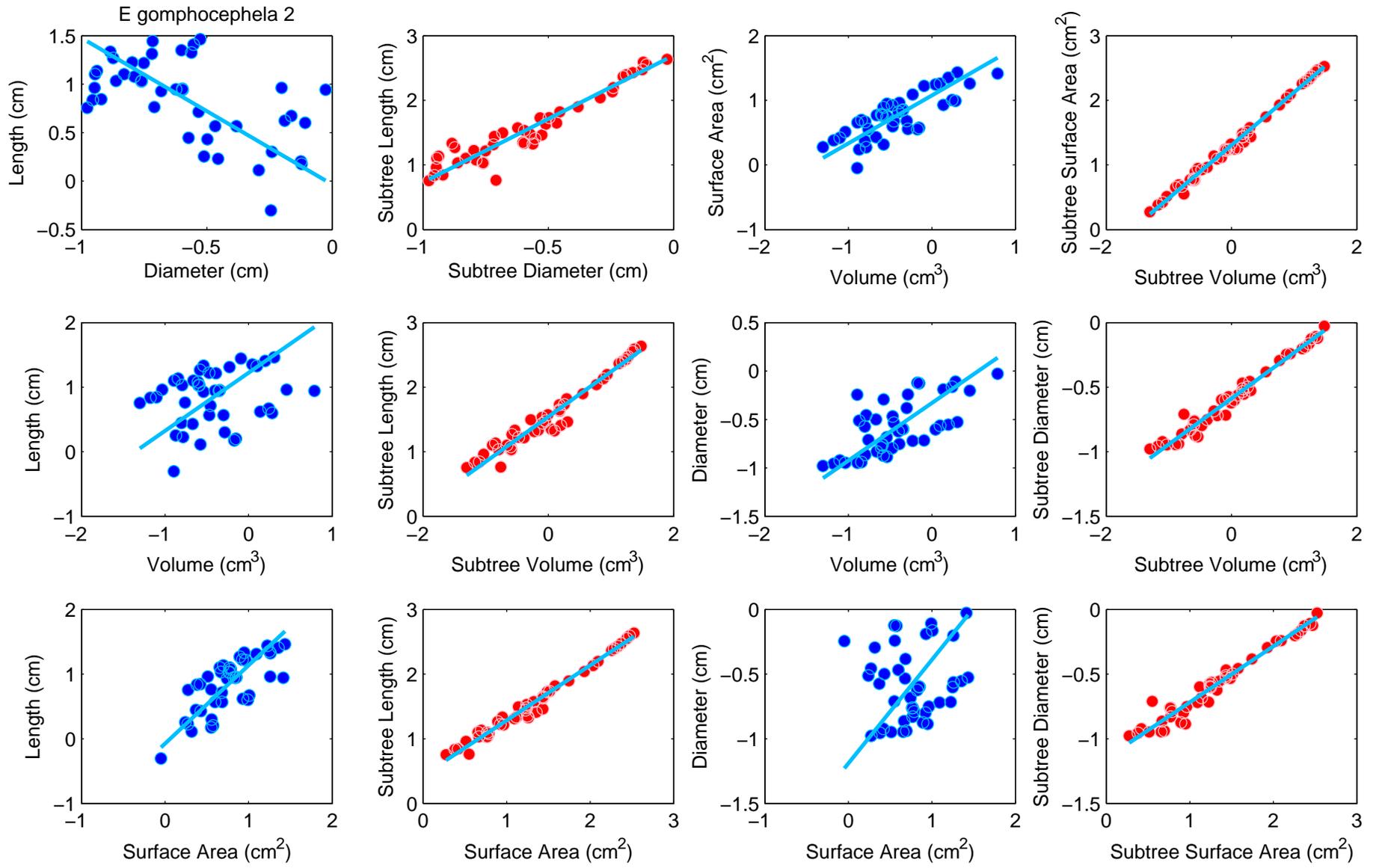

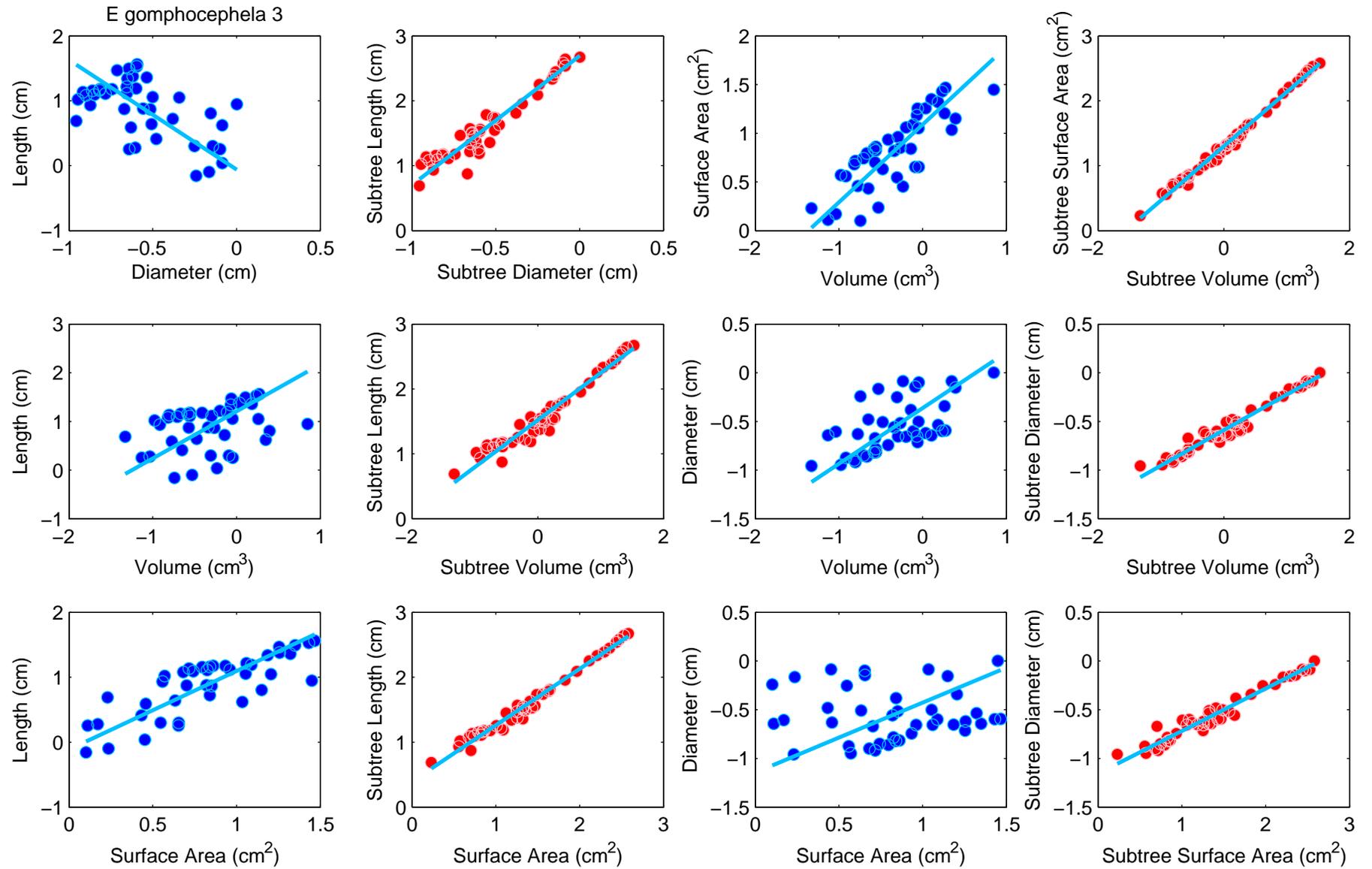

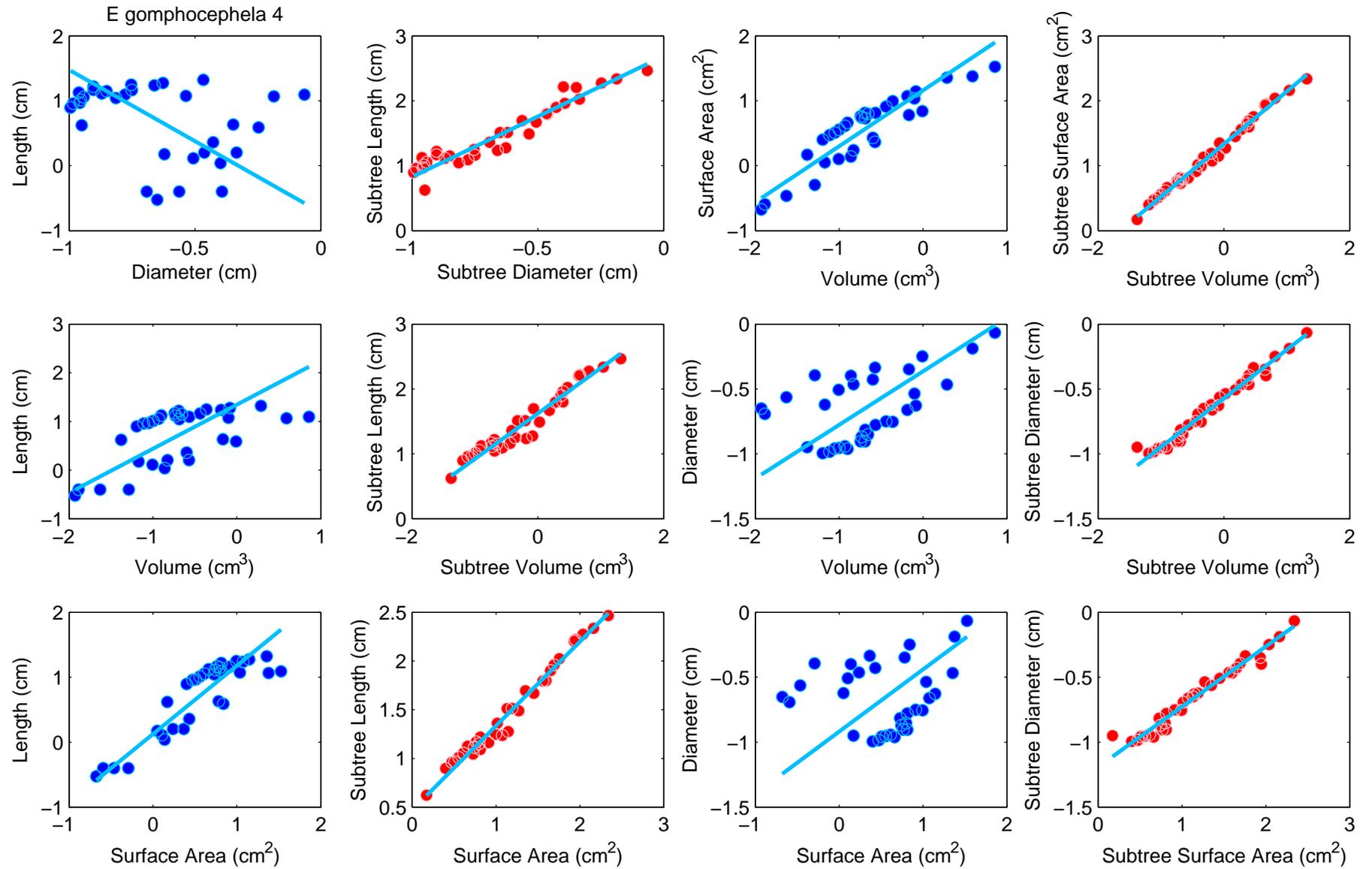

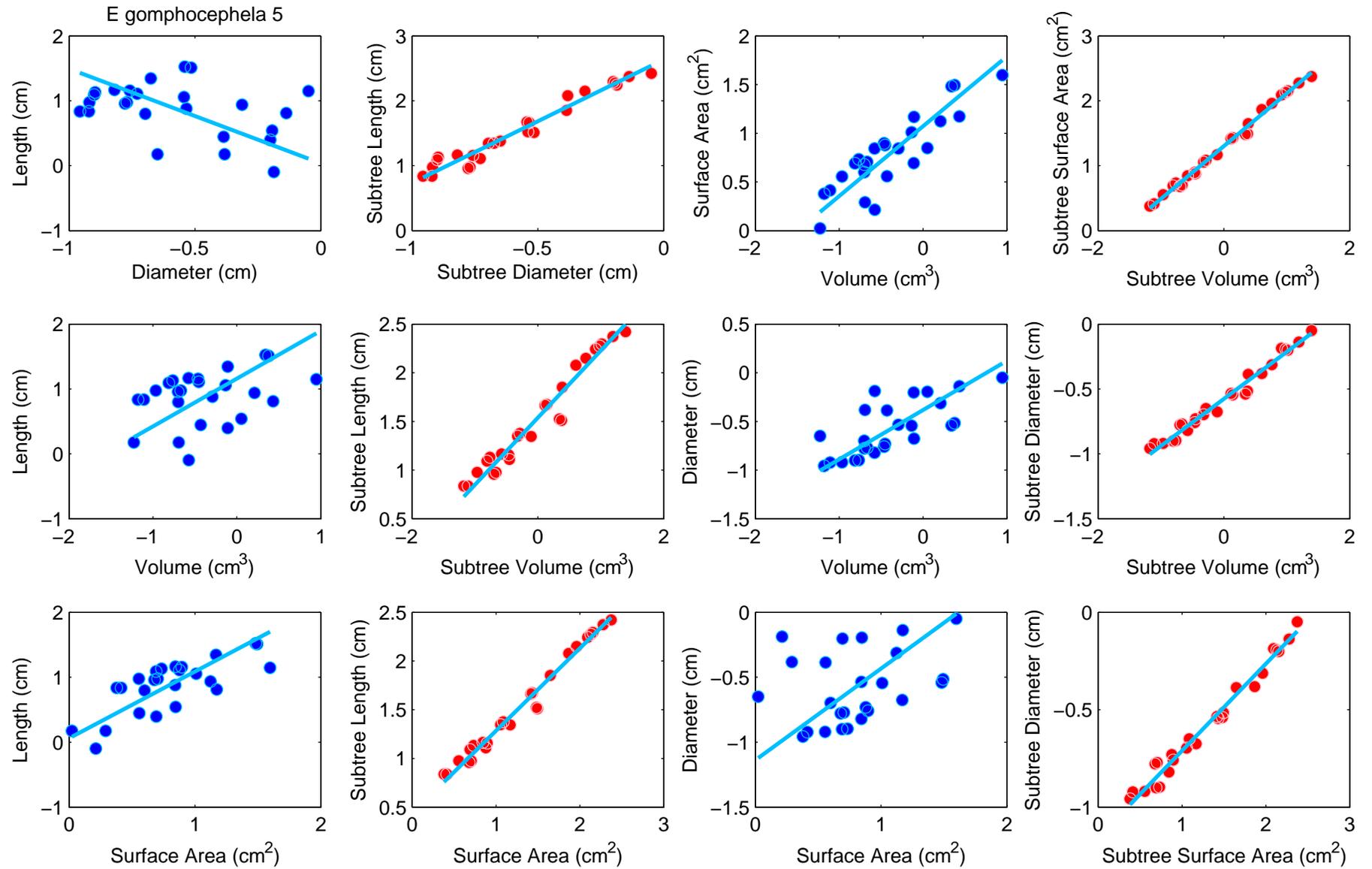

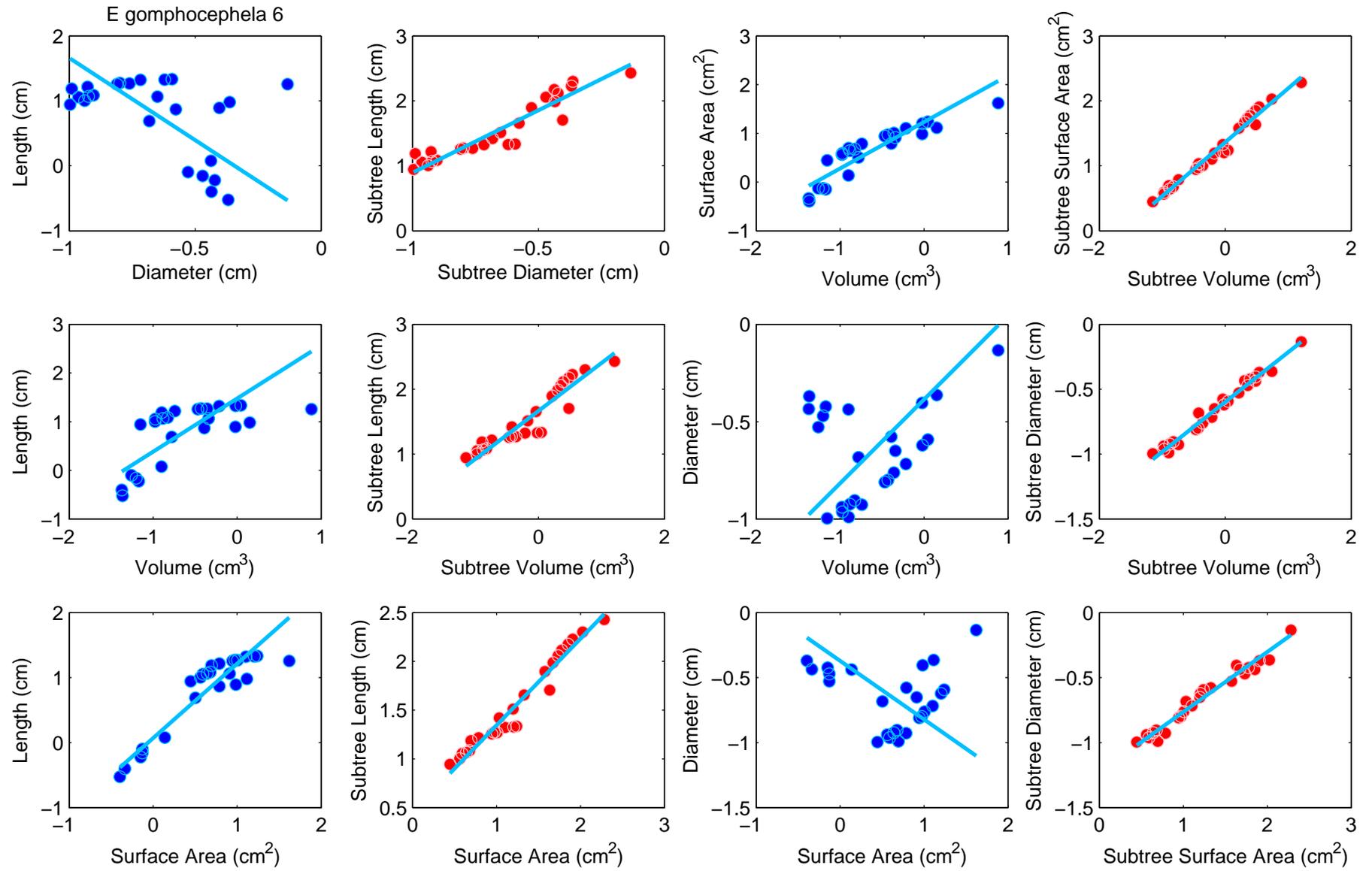

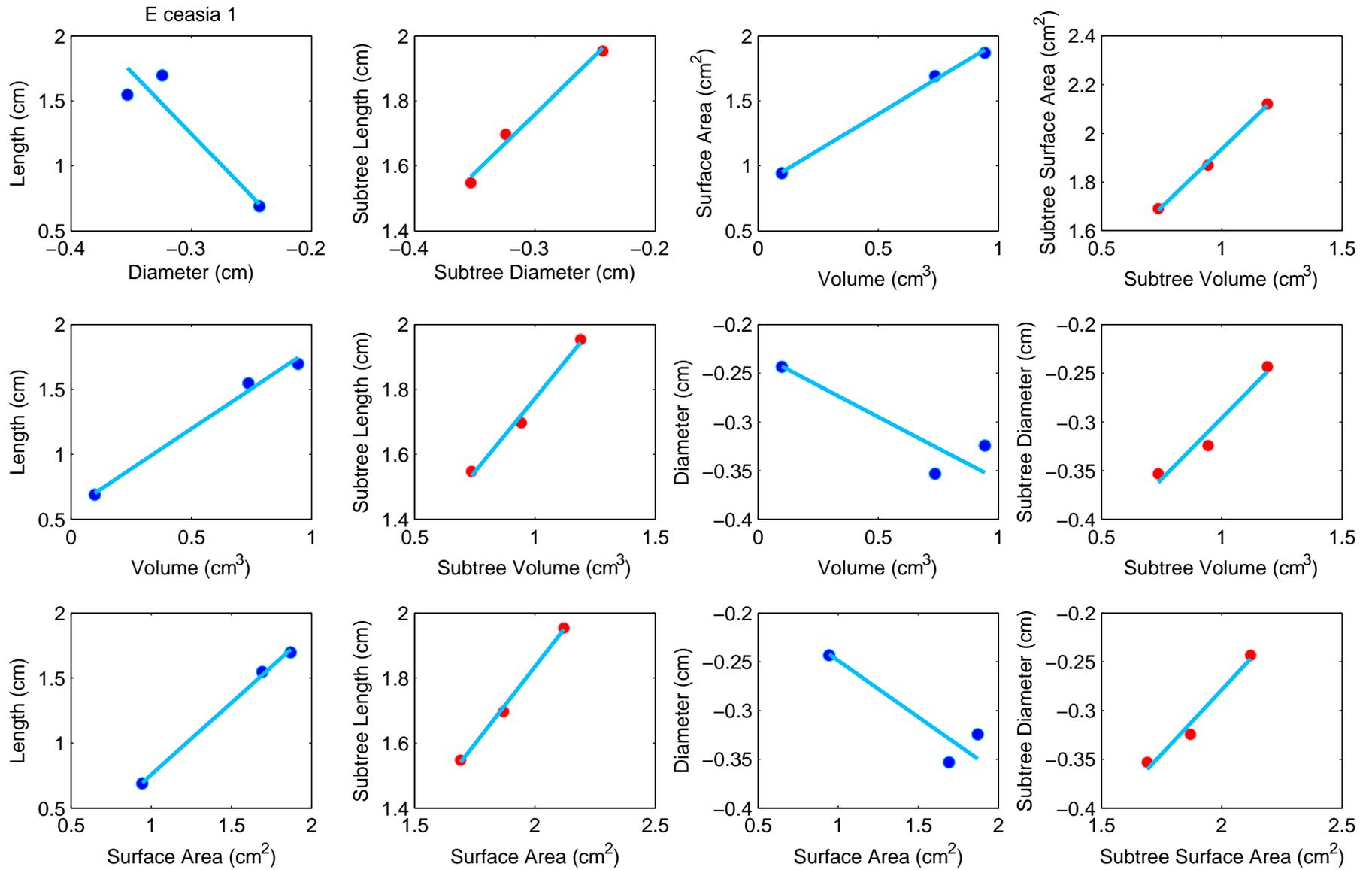

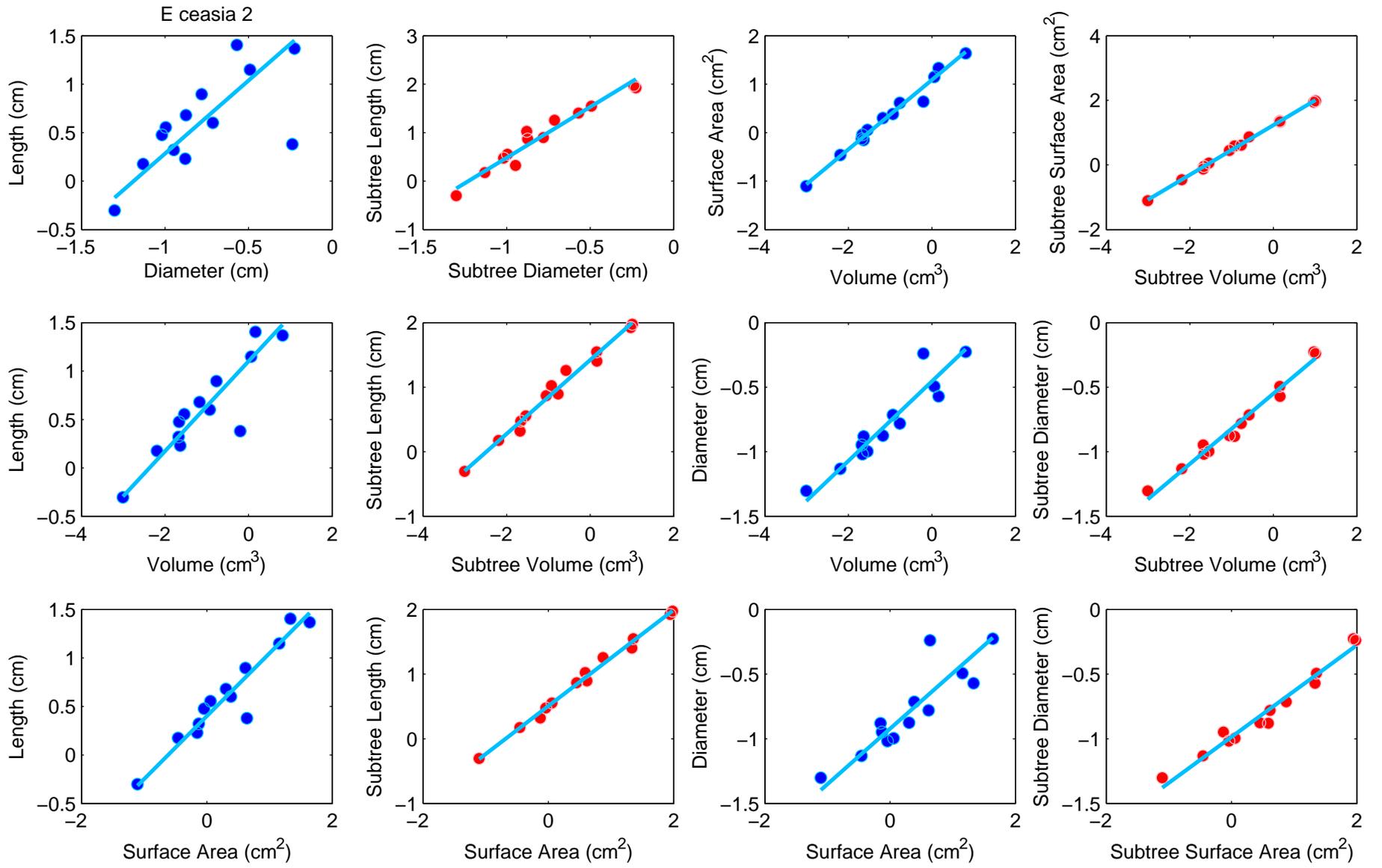

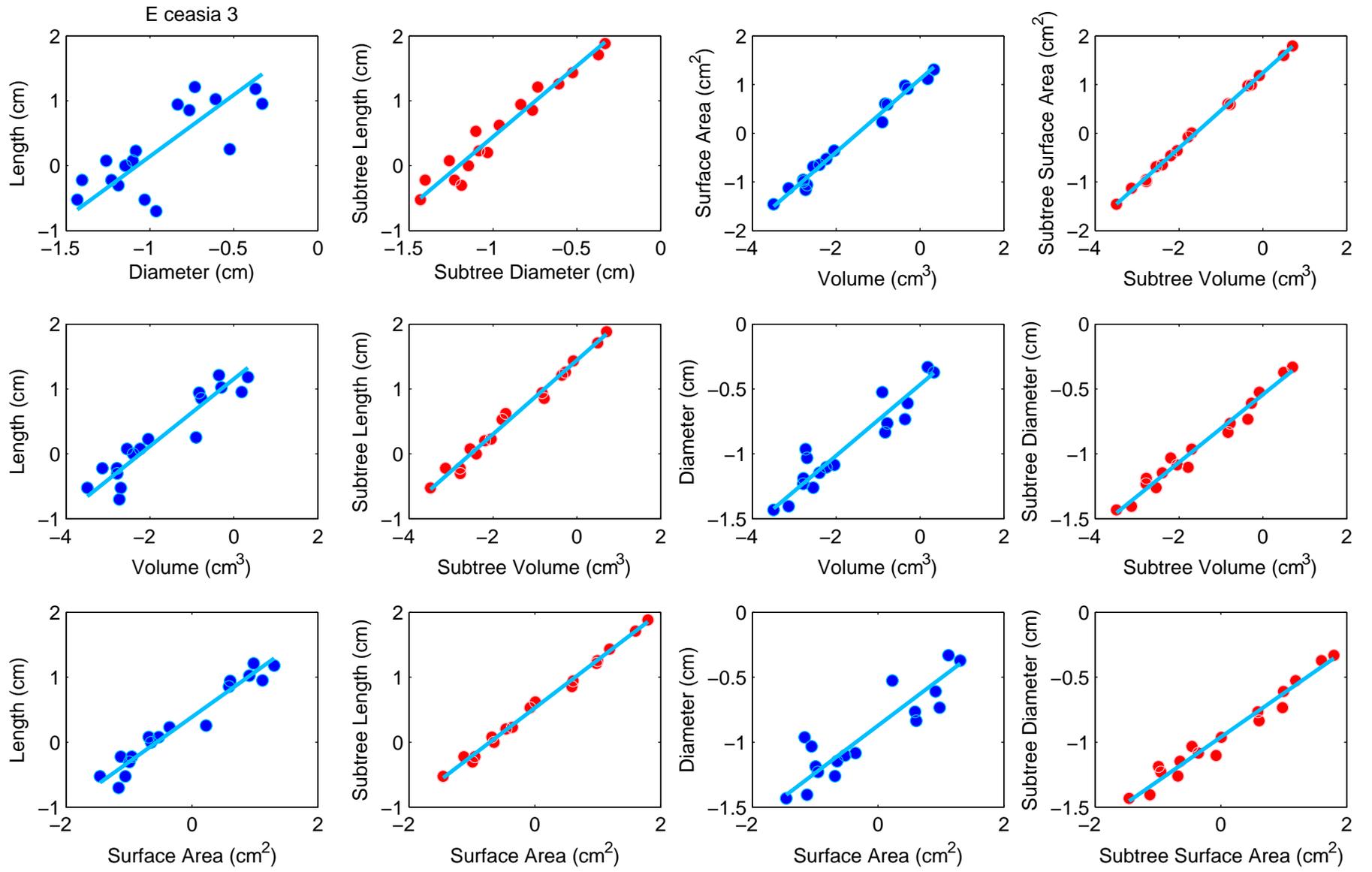

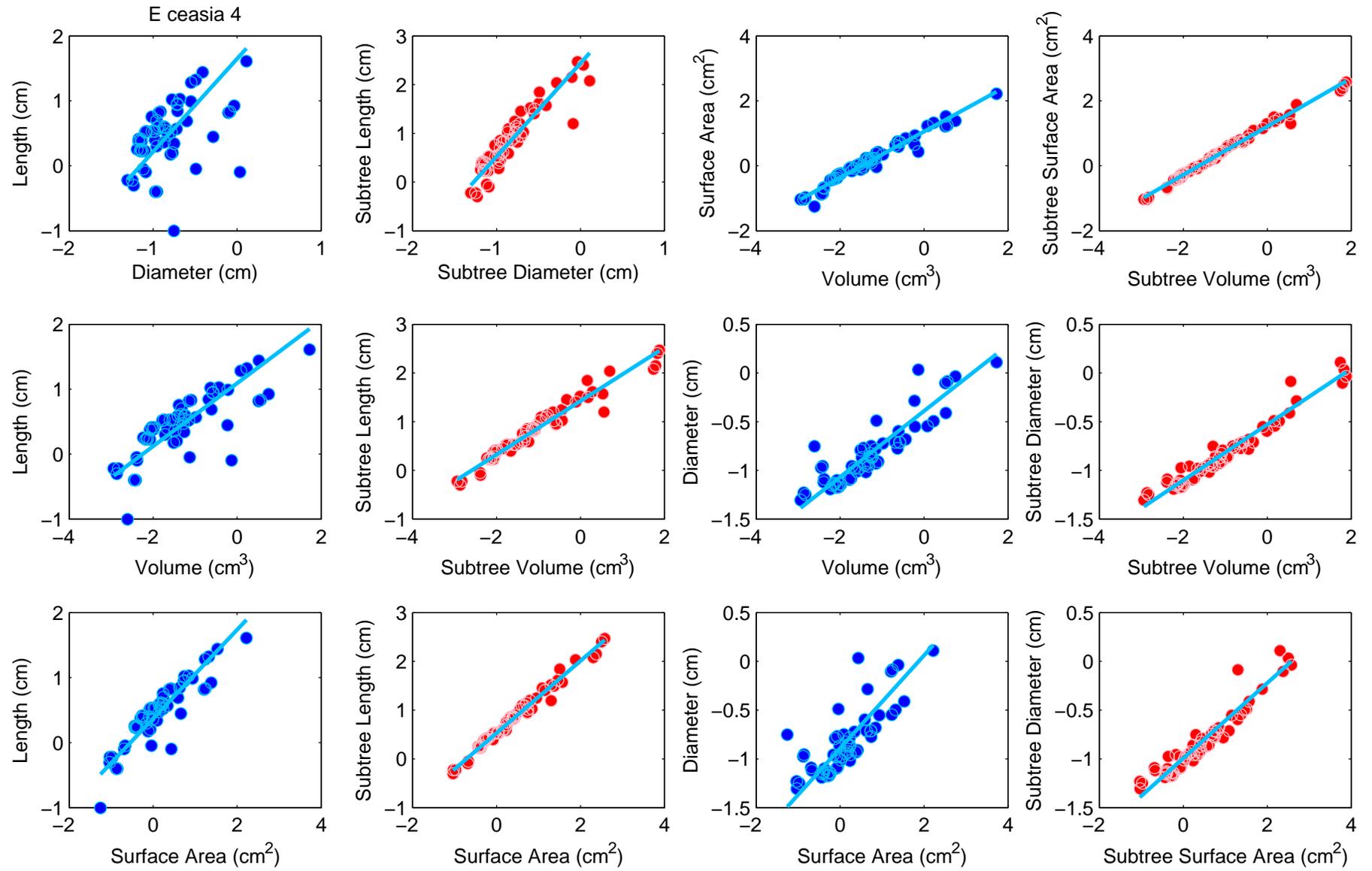

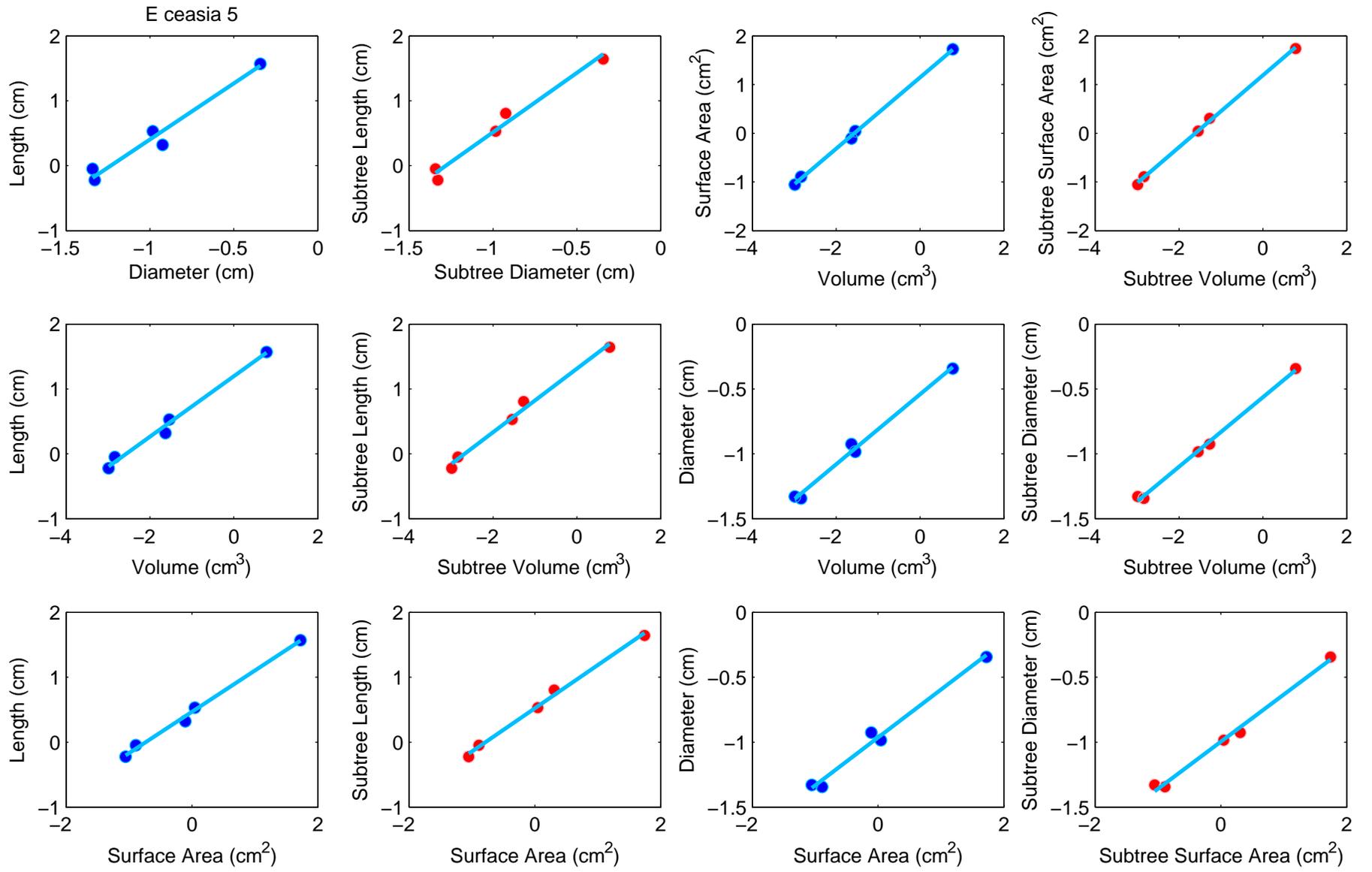

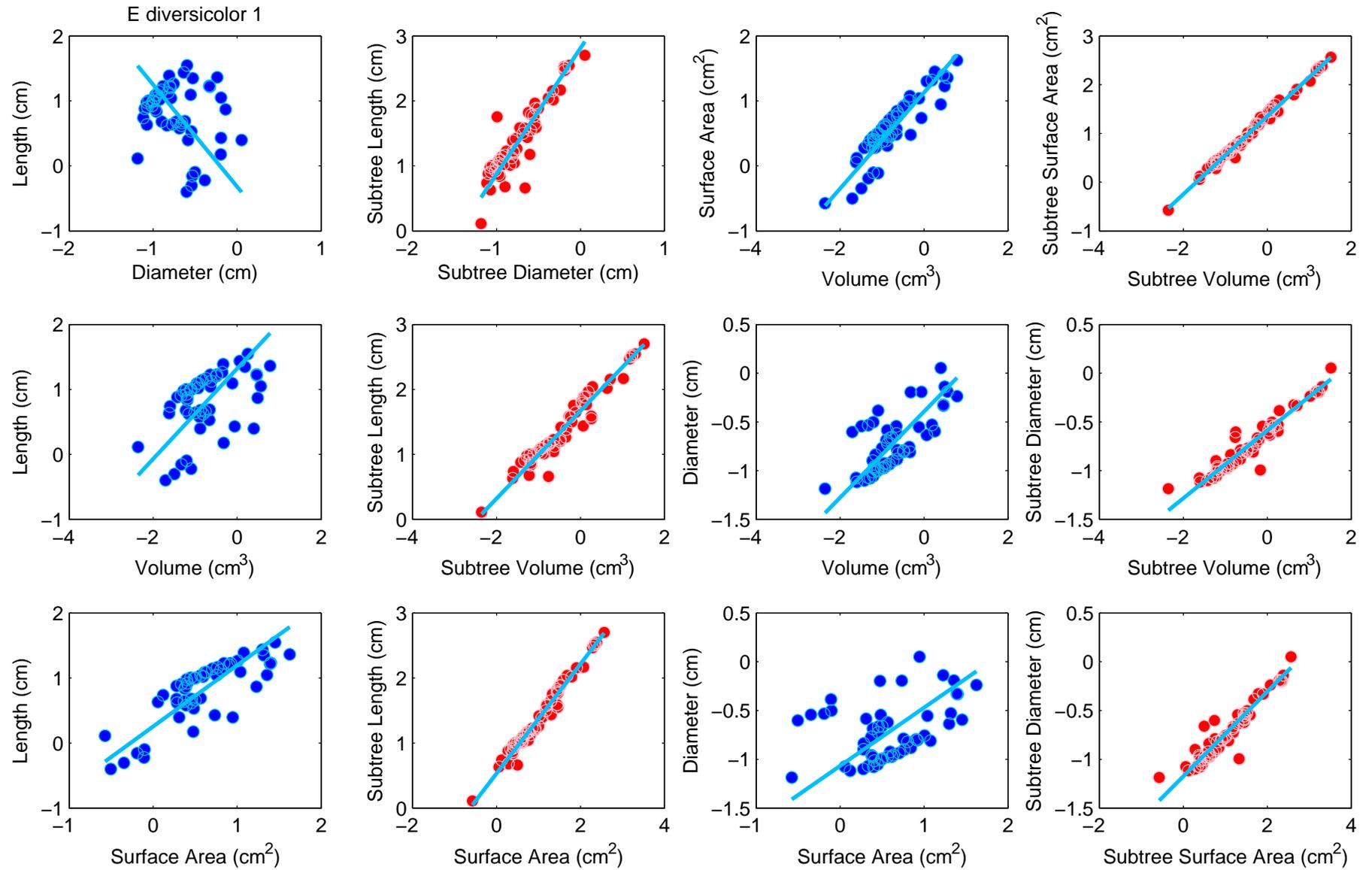

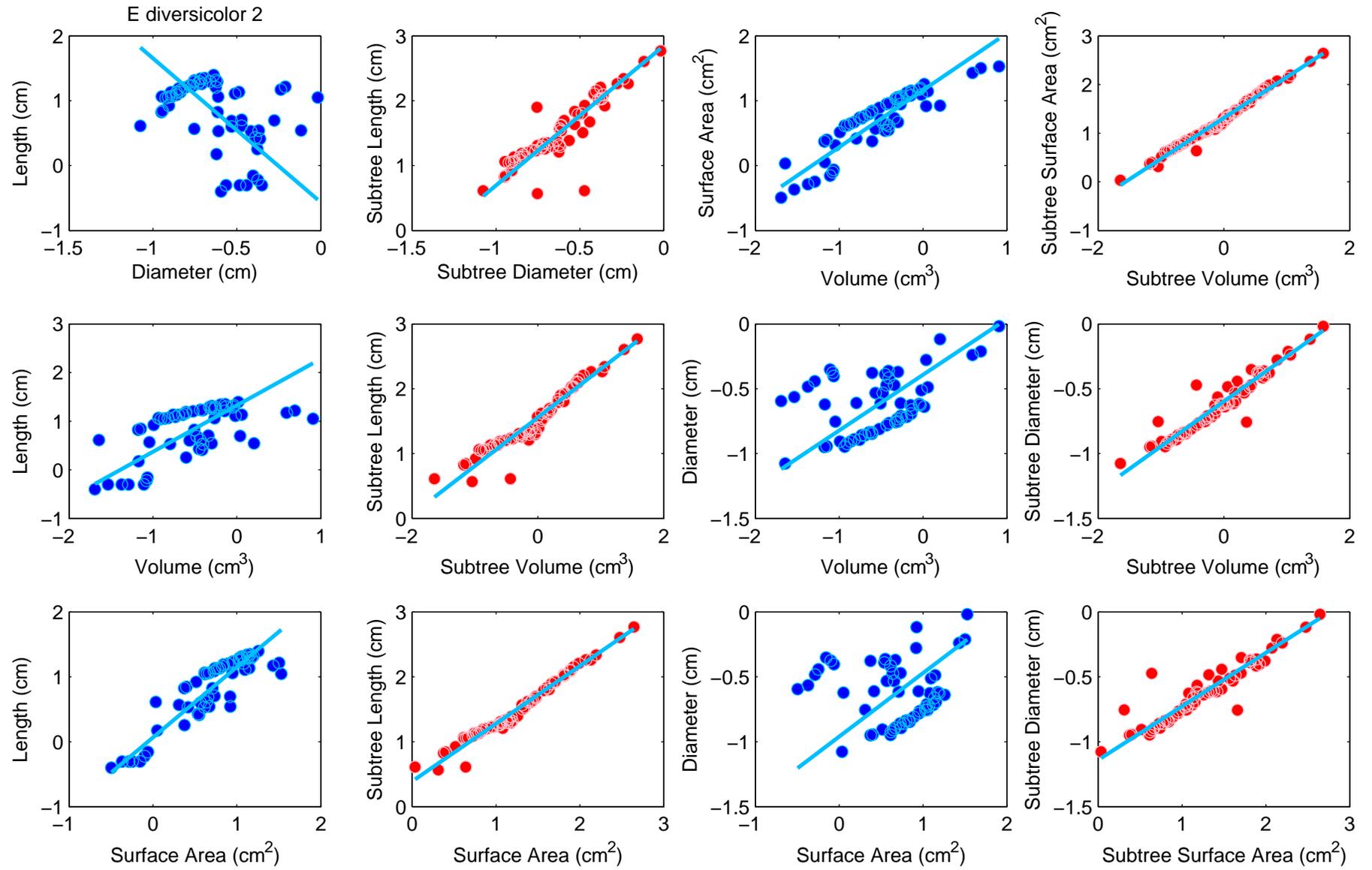

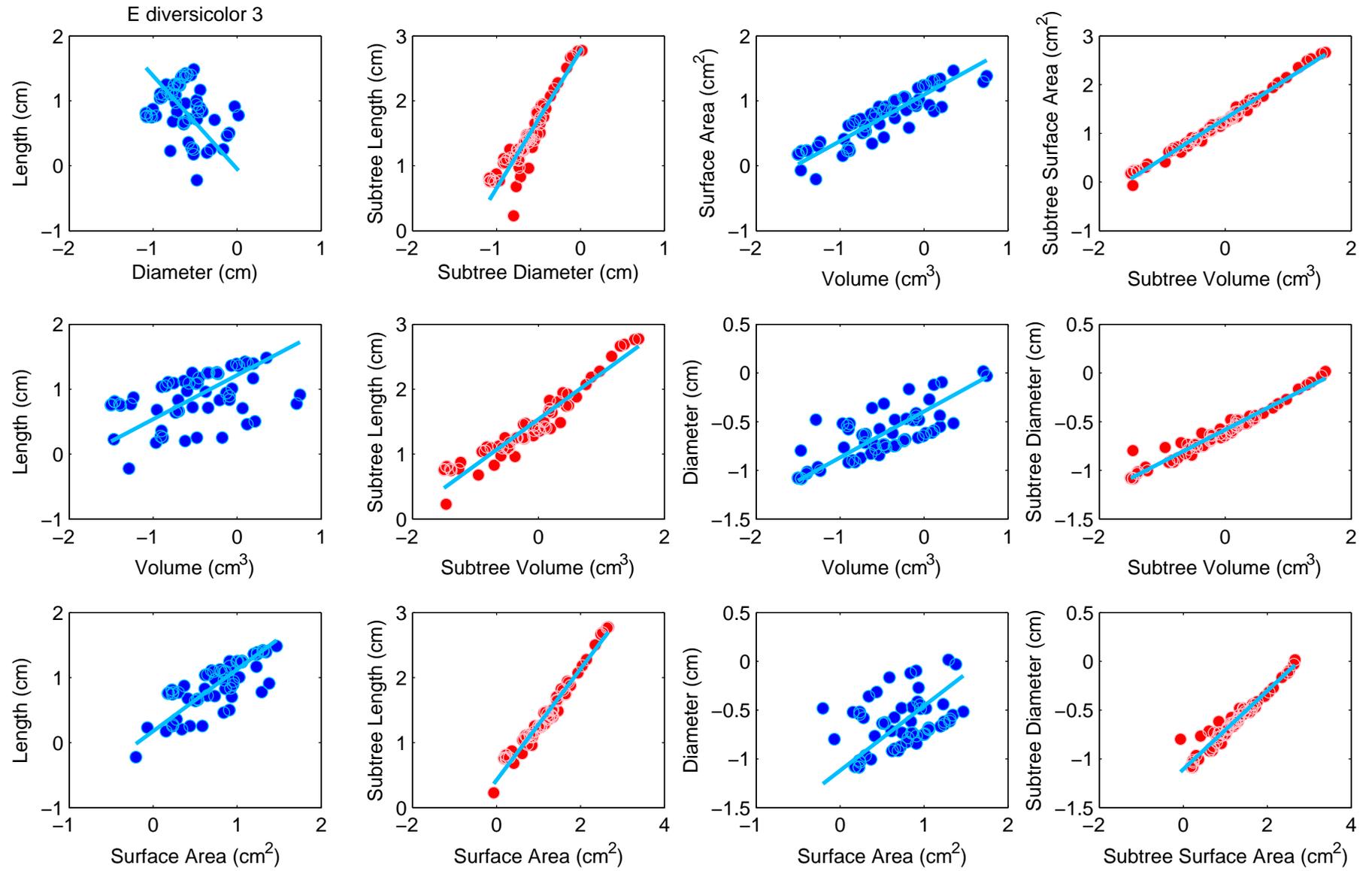

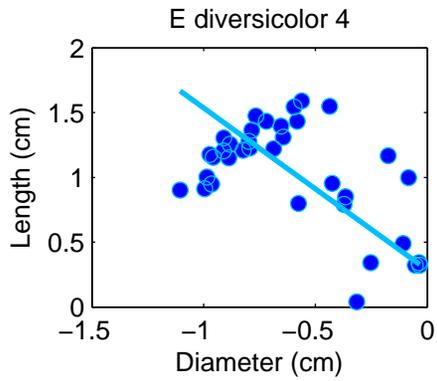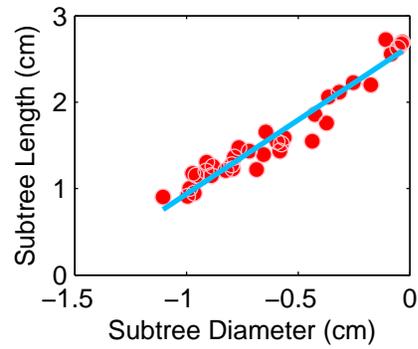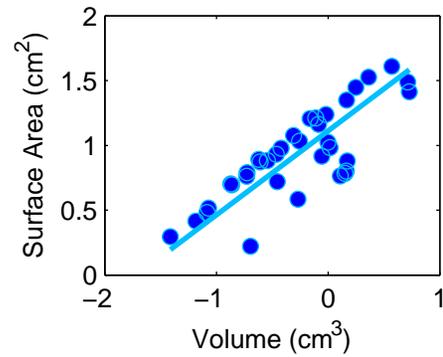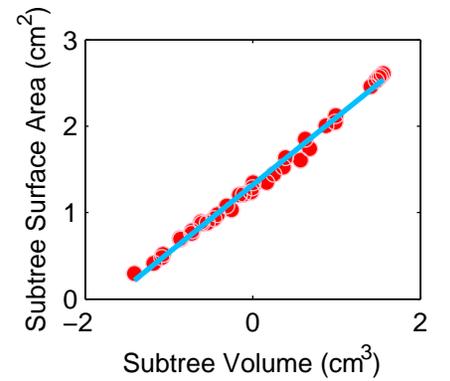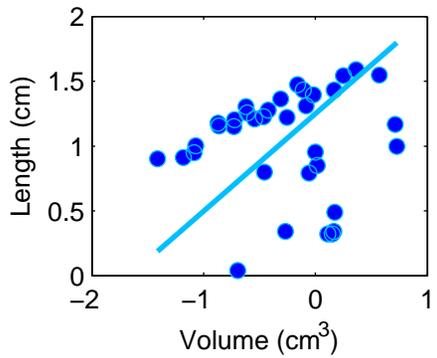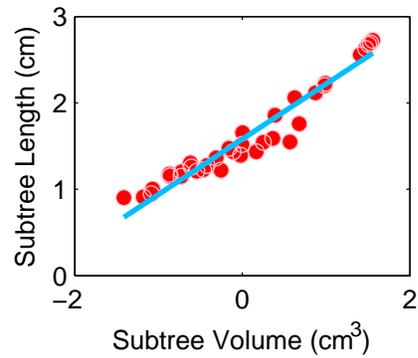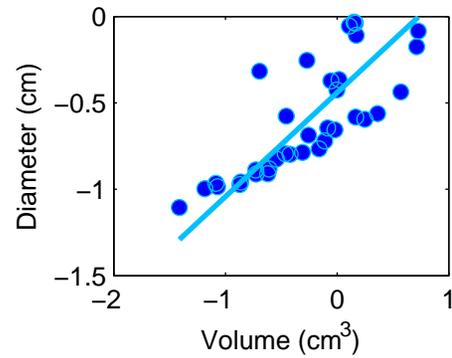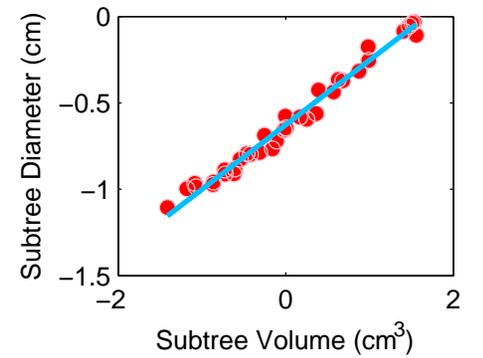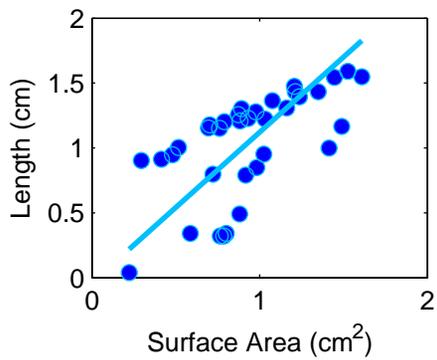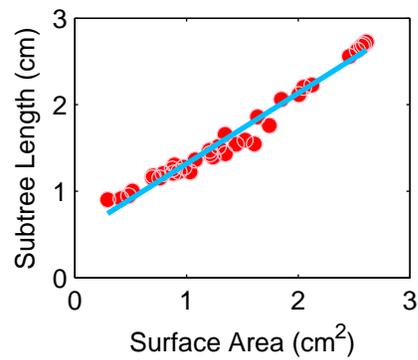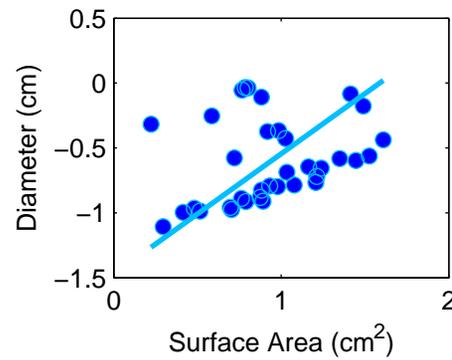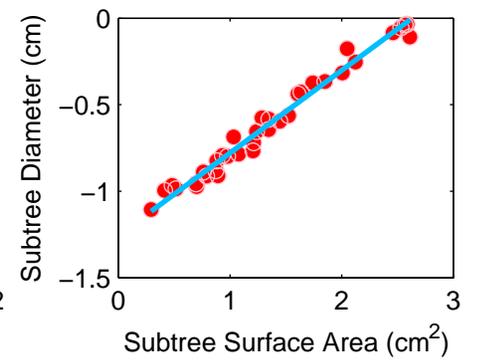

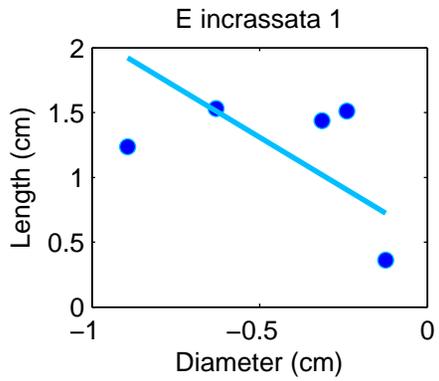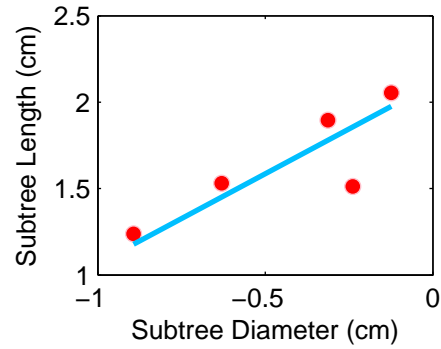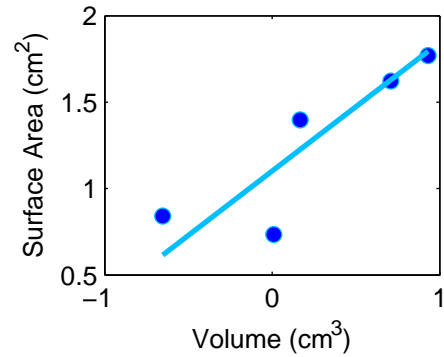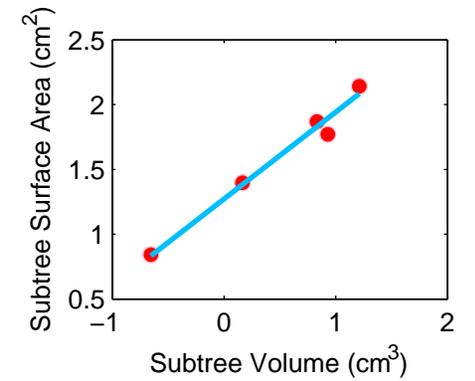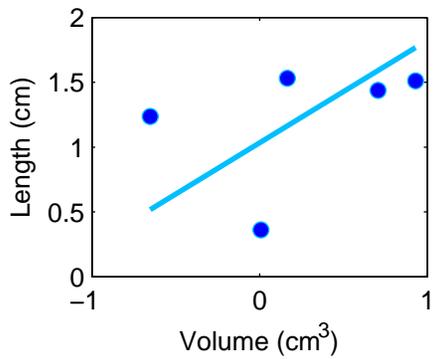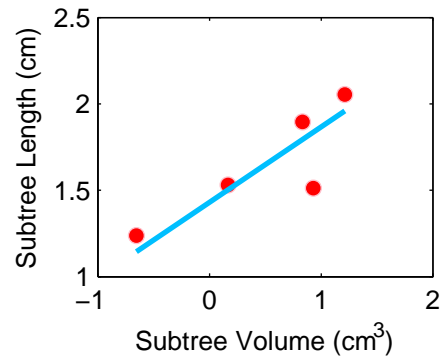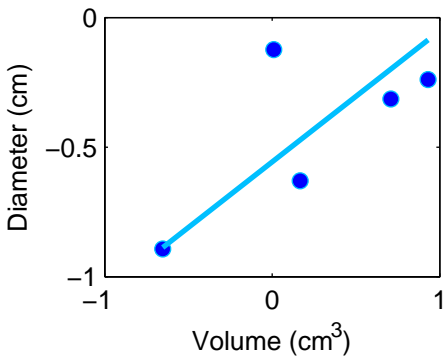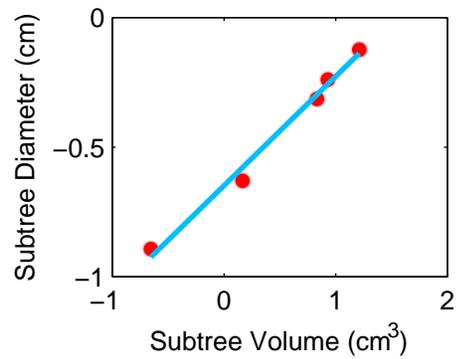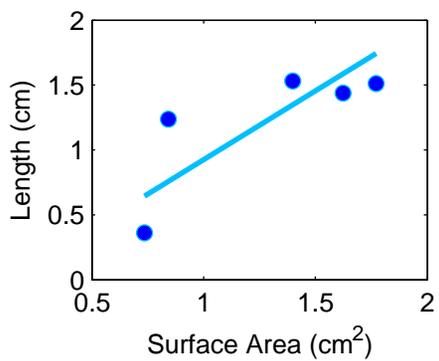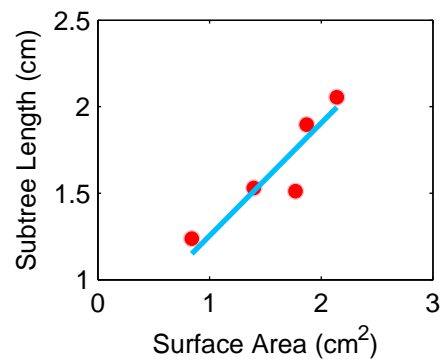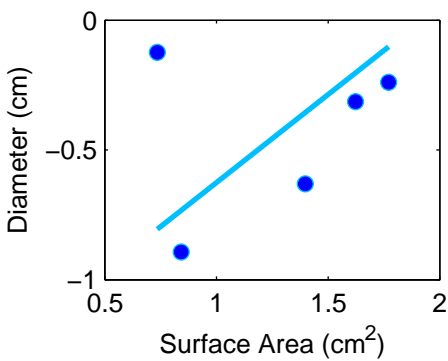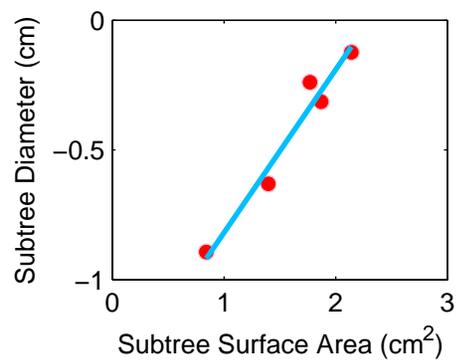

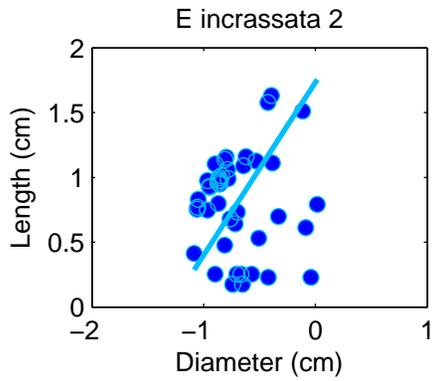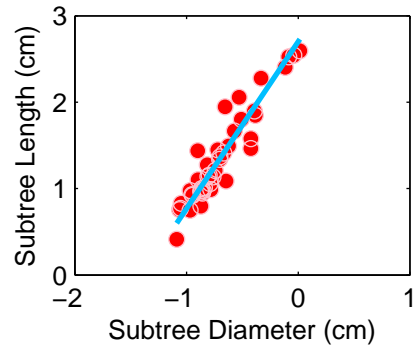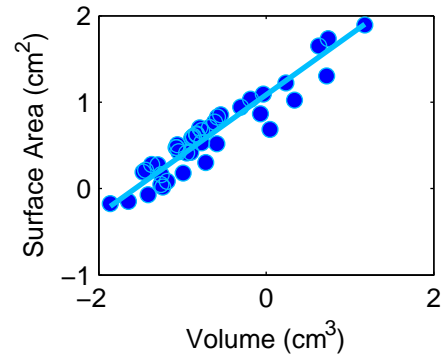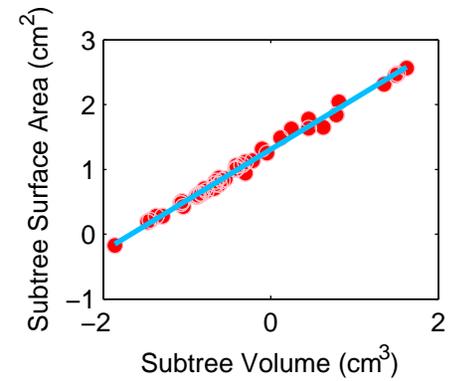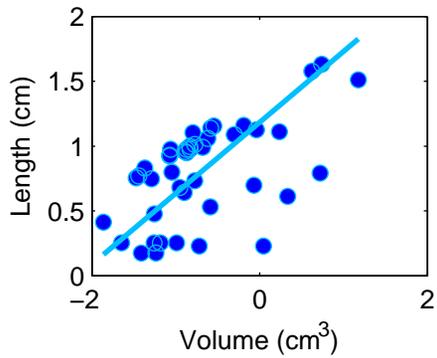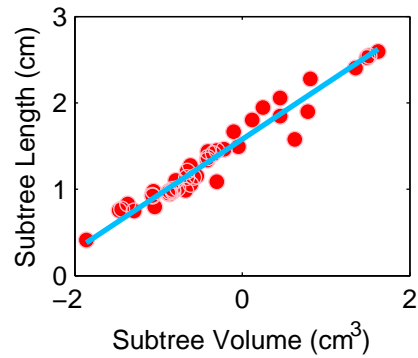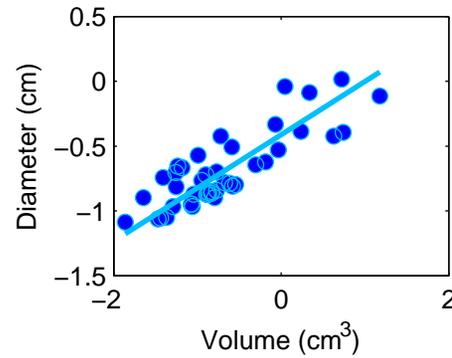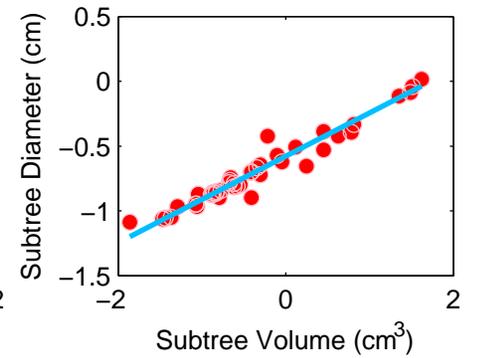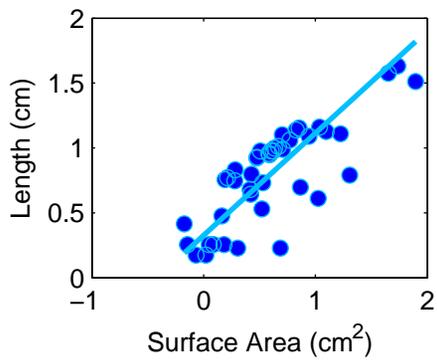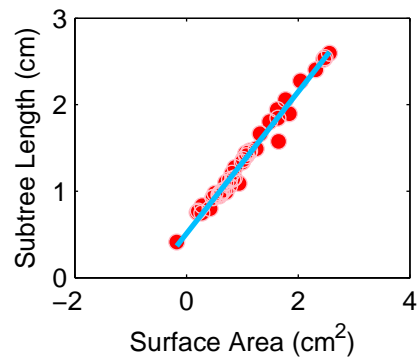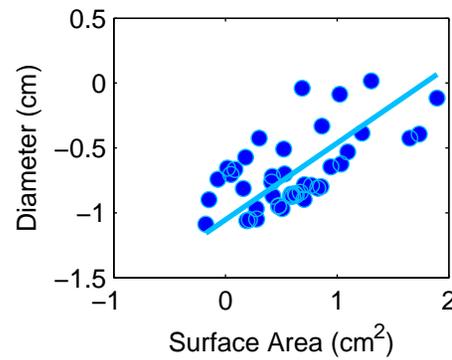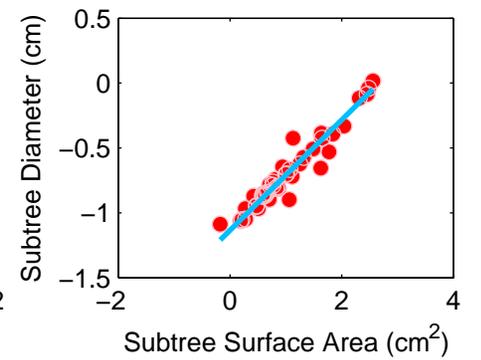

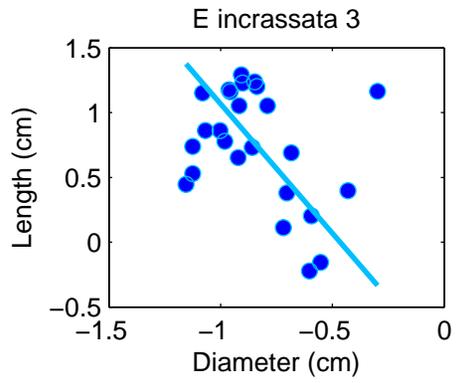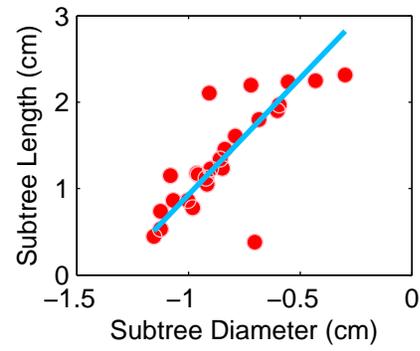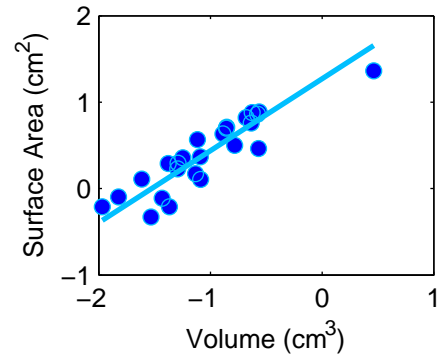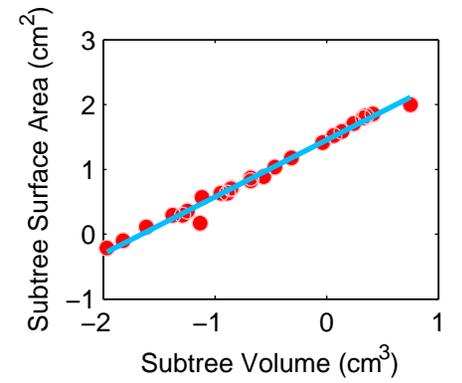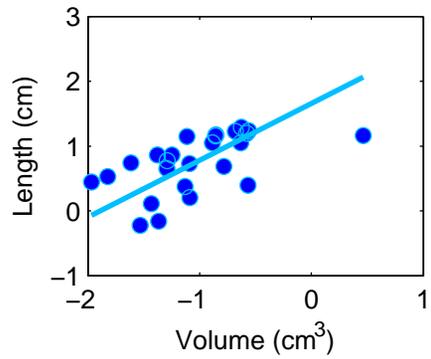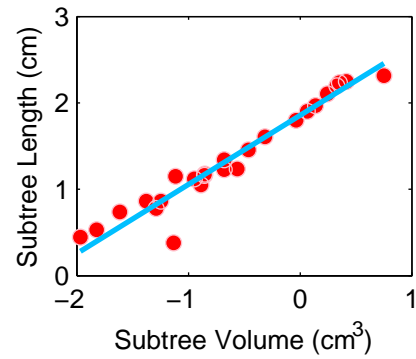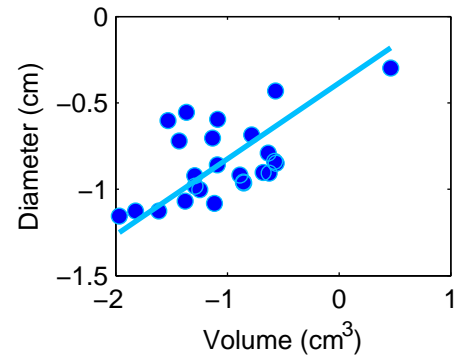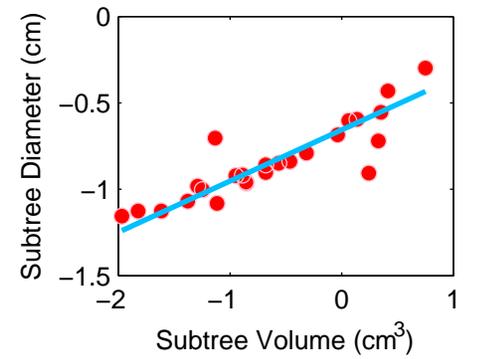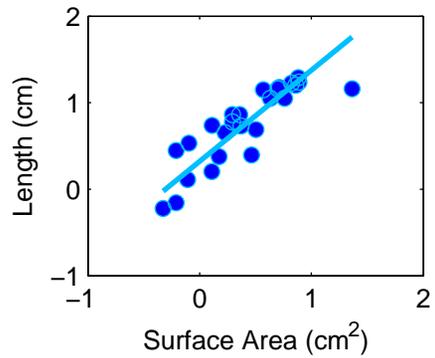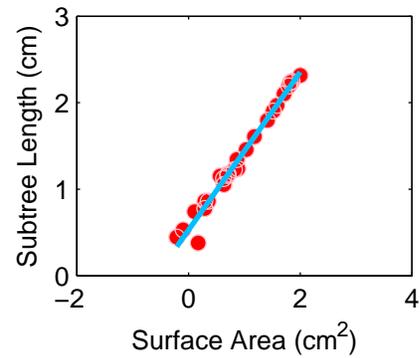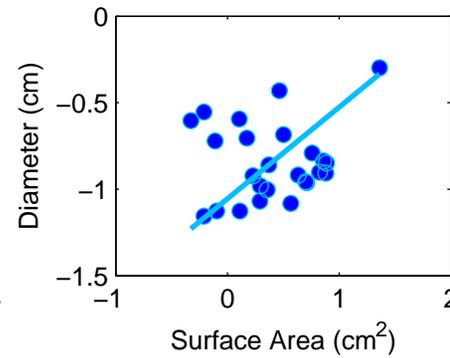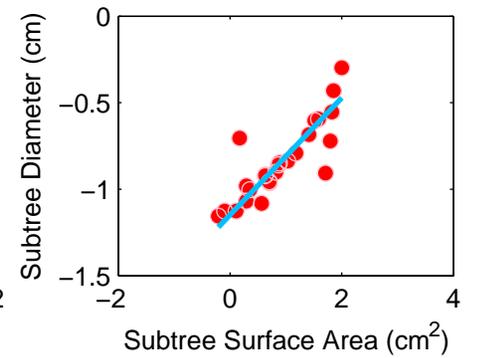

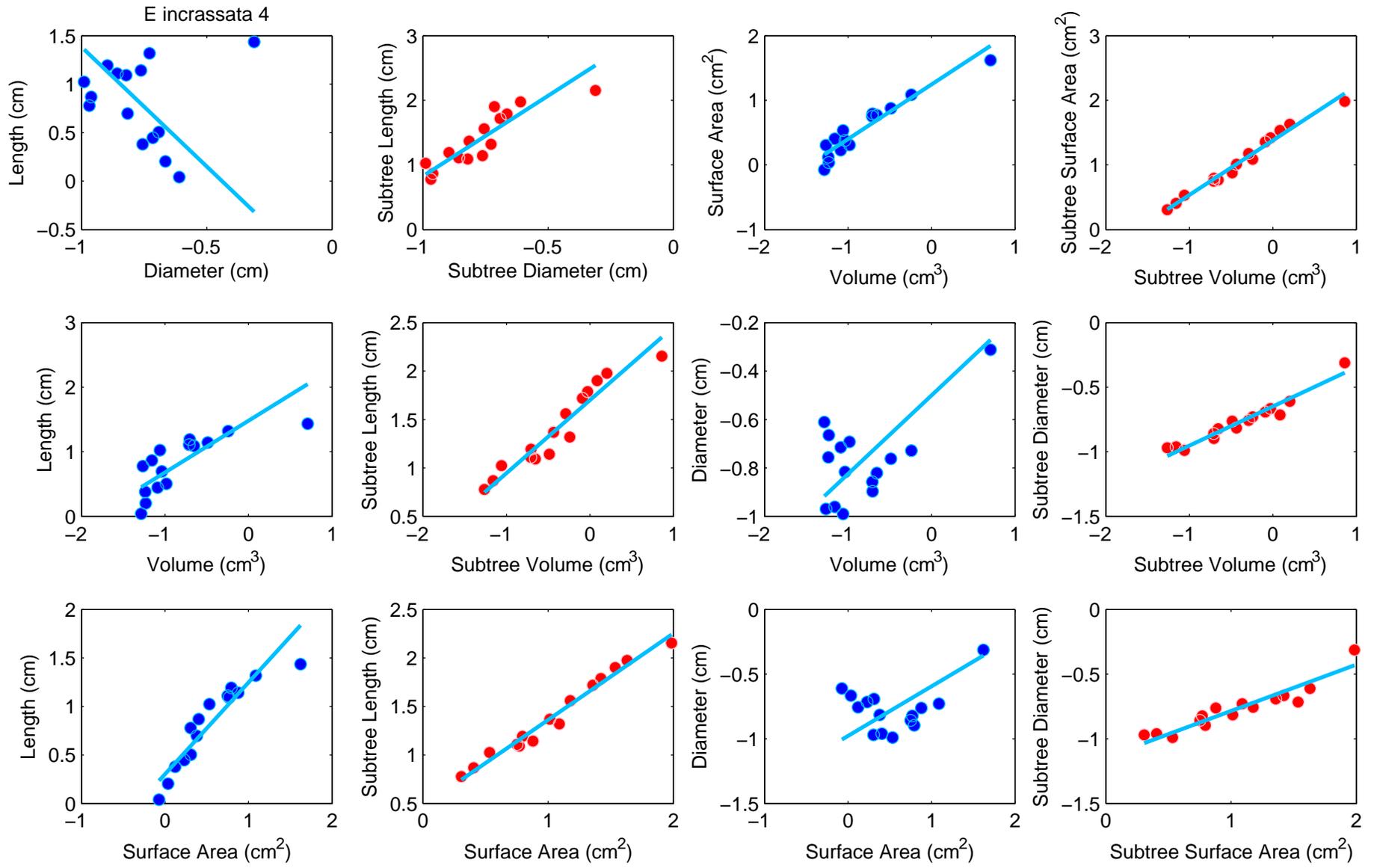

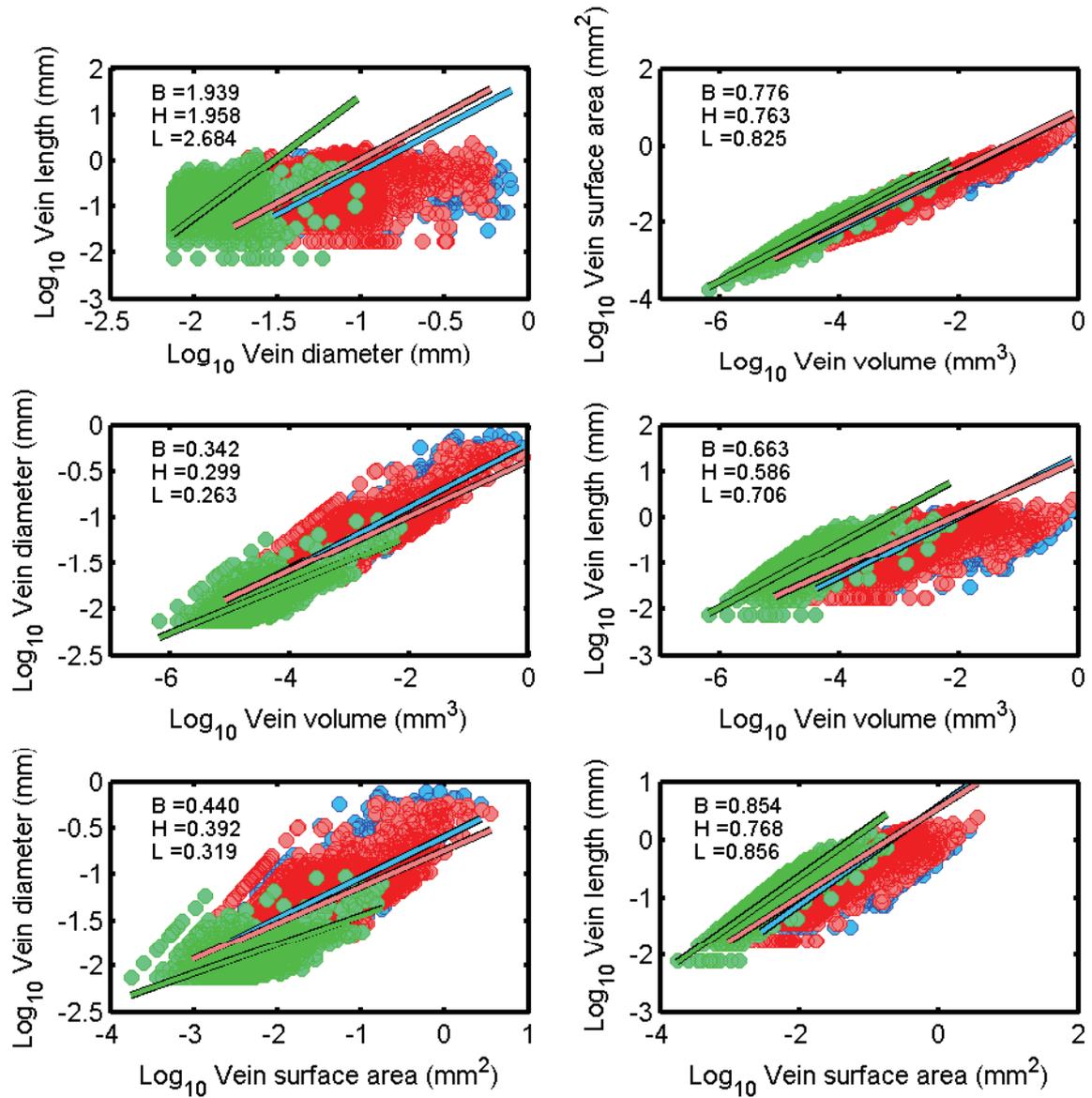

Figure S20. Bivariate relationships for the vein dimensions for the three leaves as described in the methods. The regression slope values are reported in each panel, where B, H and L correspond to *Banksia*, *Hardenbergia* and *Lespedeza*, the genus names. Full regression statistics are reported in Table S5.

**Table S1. Allometric covariation**

As shown in Table 1 of the main manuscript, the exponents for the scaling functions between length (L), diameter (D), surface area (SA) and volume (V), can all be expressed as a function of alpha ($\alpha$), the exponent from the LvD relationship. For collections of allometric exponents, one can explore the functional relationships between the scaling exponents themselves. The table below contains predictions for the 15 pairwise covariation functions between the six scaling relationships.

| Y variable | X variable | Covariation Function |
|---|---|---|
| $SA \propto V^{\frac{(\alpha+1)}{(\alpha+2)}}$ | $L \propto D^\alpha$ | $\frac{(\alpha+1)}{(\alpha+2)} = \frac{(\alpha+1)}{(\alpha+2)}$ |
| $D \propto V^{\frac{1}{(\alpha+2)}}$ | $L \propto D^\alpha$ | $\frac{1}{(\alpha+2)} = \frac{1}{(\alpha+2)}$ |
| $L \propto V^{\frac{\alpha}{(\alpha+2)}}$ | $L \propto D^\alpha$ | $\frac{\alpha}{(\alpha+2)} = \frac{\alpha}{(\alpha+2)}$ |
| $D \propto SA^{\frac{1}{(\alpha+1)}}$ | $L \propto D^\alpha$ | $\frac{1}{(\alpha+1)} = \frac{1}{(\alpha+1)}$ |
| $L \propto SA^{\frac{\alpha}{(\alpha+1)}}$ | $L \propto D^\alpha$ | $\frac{\alpha}{(\alpha+1)} = \frac{\alpha}{(\alpha+1)}$ |
| $D \propto V^{\frac{1}{(\alpha+2)}}$ | $SA \propto V^{\frac{(\alpha+1)}{(\alpha+2)}}$ | $\frac{1}{\alpha+2} = 1 - \left(\frac{\alpha+1}{\alpha+2}\right)$ |
| $L \propto V^{\frac{\alpha}{(\alpha+2)}}$ | $SA \propto V^{\frac{(\alpha+1)}{(\alpha+2)}}$ | $\frac{\alpha}{\alpha+2} = 2\left(\frac{\alpha+1}{\alpha+2}\right) - 1$ |
| $D \propto SA^{\frac{1}{(\alpha+1)}}$ | $SA \propto V^{\frac{(\alpha+1)}{(\alpha+2)}}$ | $\frac{1}{(\alpha+1)} = \frac{\alpha+2}{\alpha+1} - 1$ |
| $L \propto SA^{\frac{\alpha}{(\alpha+1)}}$ | $SA \propto V^{\frac{(\alpha+1)}{(\alpha+2)}}$ | $\frac{\alpha}{\alpha+1} = \frac{\alpha^3 + 2\alpha^2}{\alpha^3 + 3\alpha^2 + 2\alpha}$ |
| $L \propto V^{\frac{\alpha}{(\alpha+2)}}$ | $D \propto V^{\frac{1}{(\alpha+2)}}$ | $\frac{\alpha}{\alpha+2} = 1 - \left(\frac{2}{\alpha+2}\right)$ |
| $D \propto SA^{\frac{1}{(\alpha+1)}}$ | $D \propto V^{\frac{1}{(\alpha+2)}}$ | $\frac{1}{(\alpha+1)} = \frac{1}{1-\left(\frac{1}{\alpha+2}\right)} - 1$ |
| $L \propto SA^{\frac{\alpha}{(\alpha+1)}}$ | $D \propto V^{\frac{1}{(\alpha+2)}}$ | $\frac{\alpha}{\alpha+1} = \frac{1}{\left(\frac{1}{\alpha+2}\right)-1} + 2$ |
| $D \propto SA^{\frac{1}{(\alpha+1)}}$ | $L \propto V^{\frac{\alpha}{(\alpha+2)}}$ | $\frac{1}{(\alpha+1)} = \frac{2}{\left(\frac{\alpha}{\alpha+2}\right)+1} - 1$ |
| $L \propto SA^{\frac{\alpha}{(\alpha+1)}}$ | $L \propto V^{\frac{\alpha}{(\alpha+2)}}$ | $\frac{\alpha}{(\alpha+1)} = \frac{\alpha^2 + 2\alpha}{\alpha^2 + 3\alpha + 2}$ |
| $D \propto SA^{\frac{1}{(\alpha+1)}}$ | $L \propto SA^{\frac{\alpha}{(\alpha+1)}}$ | $\frac{1}{\alpha+1} = 1 - \left(\frac{\alpha}{\alpha+1}\right)$ |

**Table S2.** SMA regression data for the 19 individual saplings from 4 species within the Eucalyptus genus. Data are grouped by relationship (first two columns) and whether the regressions were done on raw data or subtrees as described in the methods.

| Y data | X data | Group | Species | Individual | n | R2 | P | Slope | LowCI | UppCI | Interc | LowCI | UppCI | Ymean | Xmean |
|---|---|---|---|---|---|---|---|---|---|---|---|---|---|---|---|
| Log10 Stem Length (cm) | Log10 Stem Diameter (cm) | Raw data | E. gomphocephela | 1 | 36 | 0 | 0.925 | -1.098 | -1.546 | -0.78 | -0.12694 | -0.45365 | 0.19978 | 0.691 | -0.745 |
| Log10 Stem Length (cm) | Log10 Stem Diameter (cm) | Raw data | E. gomphocephela | 2 | 43 | 0.327 | 0 | -1.513 | -1.954 | -1.171 | -0.03418 | -0.29098 | 0.22261 | 0.845 | -0.581 |
| Log10 Stem Length (cm) | Log10 Stem Diameter (cm) | Raw data | E. gomphocephela | 3 | 41 | 0.315 | 0 | -1.687 | -2.199 | -1.294 | -0.05673 | -0.34134 | 0.22787 | 0.88 | -0.555 |
| Log10 Stem Length (cm) | Log10 Stem Diameter (cm) | Raw data | E. gomphocephela | 4 | 33 | 0.114 | 0.055 | -2.198 | -3.083 | -1.567 | -0.72098 | -1.26812 | -0.17384 | 0.713 | -0.653 |
| Log10 Stem Length (cm) | Log10 Stem Diameter (cm) | Raw data | E. gomphocephela | 5 | 25 | 0.146 | 0.059 | -1.458 | -2.15 | -0.988 | 0.0346 | -0.34982 | 0.41902 | 0.877 | -0.578 |
| Log10 Stem Length (cm) | Log10 Stem Diameter (cm) | Raw data | E. gomphocephela | 6 | 25 | 0.247 | 0.011 | -2.537 | -3.658 | -1.759 | -0.87288 | -1.54528 | -0.20048 | 0.792 | -0.656 |
| Log10 Stem Length (cm) | Log10 Stem Diameter (cm) | Raw data | E. ceasia | 1 | 13 | 0.504 | 0.007 | 1.509 | 0.961 | 2.371 | 1.79121 | 1.19431 | 2.38812 | 0.611 | -0.782 |
| Log10 Stem Length (cm) | Log10 Stem Diameter (cm) | Raw data | E. ceasia | 2 | 17 | 0.582 | 0 | 1.898 | 1.339 | 2.689 | 2.0414 | 1.36545 | 2.71735 | 0.255 | -0.941 |
| Log10 Stem Length (cm) | Log10 Stem Diameter (cm) | Raw data | E. ceasia | 3 | 56 | 0.218 | 0 | 1.432 | 1.128 | 1.819 | 1.64993 | 1.33617 | 1.96369 | 0.476 | -0.82 |
| Log10 Stem Length (cm) | Log10 Stem Diameter (cm) | Raw data | E. diversicolor | 1 | 52 | 0.046 | 0.128 | -1.574 | -2.071 | -1.197 | -0.32131 | -0.67474 | 0.03211 | 0.809 | -0.718 |
| Log10 Stem Length (cm) | Log10 Stem Diameter (cm) | Raw data | E. diversicolor | 2 | 57 | 0.157 | 0.002 | -2.22 | -2.838 | -1.736 | -0.56542 | -0.93965 | -0.19118 | 0.816 | -0.622 |
| Log10 Stem Length (cm) | Log10 Stem Diameter (cm) | Raw data | E. diversicolor | 3 | 59 | 0.076 | 0.034 | -1.438 | -1.851 | -1.118 | -0.049 | -0.3116 | 0.2136 | 0.869 | -0.638 |
| Log10 Stem Length (cm) | Log10 Stem Diameter (cm) | Raw data | E. diversicolor | 4 | 35 | 0.314 | 0 | -1.249 | -1.668 | -0.935 | 0.28588 | 0.02875 | 0.54302 | 1.043 | -0.606 |
| Log10 Stem Length (cm) | Log10 Stem Diameter (cm) | Raw data | E. incrassata | 1 | 43 | 0.001 | 0.849 | 1.333 | 0.977 | 1.818 | 1.73285 | 1.39923 | 2.06647 | 0.807 | -0.695 |
| Log10 Stem Length (cm) | Log10 Stem Diameter (cm) | Raw data | E. incrassata | 2 | 25 | 0.124 | 0.085 | -1.997 | -2.96 | -1.347 | -0.92929 | -1.63879 | -0.21979 | 0.75 | -0.841 |
| Log10 Stem Length (cm) | Log10 Stem Diameter (cm) | Raw data | E. incrassata | 3 | 15 | 0.004 | 0.829 | -2.473 | -4.36 | -1.402 | -1.08685 | -2.27026 | 0.09655 | 0.816 | -0.77 |
| Log10 Stem Length (cm) | Log10 Stem Diameter (cm) | Subtree data | E. gomphocephela | 1 | 36 | 0.941 | 0 | 1.907 | 1.752 | 2.076 | 2.62854 | 2.49869 | 2.75838 | 1.208 | -0.745 |
| Log10 Stem Length (cm) | Log10 Stem Diameter (cm) | Subtree data | E. gomphocephela | 2 | 43 | 0.912 | 0 | 1.966 | 1.79 | 2.159 | 2.69559 | 2.57719 | 2.81398 | 1.553 | -0.581 |
| Log10 Stem Length (cm) | Log10 Stem Diameter (cm) | Subtree data | E. gomphocephela | 3 | 41 | 0.919 | 0 | 1.986 | 1.812 | 2.178 | 2.69105 | 2.57859 | 2.80351 | 1.588 | -0.555 |
| Log10 Stem Length (cm) | Log10 Stem Diameter (cm) | Subtree data | E. gomphocephela | 4 | 33 | 0.923 | 0 | 1.864 | 1.684 | 2.063 | 2.68948 | 2.55679 | 2.82218 | 1.473 | -0.653 |
| Log10 Stem Length (cm) | Log10 Stem Diameter (cm) | Subtree data | E. gomphocephela | 5 | 25 | 0.959 | 0 | 1.901 | 1.741 | 2.075 | 2.63411 | 2.52783 | 2.74038 | 1.536 | -0.578 |
| Log10 Stem Length (cm) | Log10 Stem Diameter (cm) | Subtree data | E. gomphocephela | 6 | 25 | 0.883 | 0 | 1.924 | 1.661 | 2.229 | 2.81667 | 2.6189 | 3.01443 | 1.554 | -0.656 |
| Log10 Stem Length (cm) | Log10 Stem Diameter (cm) | Subtree data | E. ceasia | 1 | 13 | 0.951 | 0 | 2.097 | 1.81 | 2.428 | 2.57291 | 2.31423 | 2.8316 | 0.934 | -0.782 |
| Log10 Stem Length (cm) | Log10 Stem Diameter (cm) | Subtree data | E. ceasia | 2 | 17 | 0.949 | 0 | 2.187 | 1.932 | 2.476 | 2.62966 | 2.35882 | 2.9005 | 0.571 | -0.941 |
| Log10 Stem Length (cm) | Log10 Stem Diameter (cm) | Subtree data | E. ceasia | 3 | 56 | 0.859 | 0 | 1.918 | 1.732 | 2.125 | 2.43643 | 2.26185 | 2.61101 | 0.864 | -0.82 |
| Log10 Stem Length (cm) | Log10 Stem Diameter (cm) | Subtree data | E. diversicolor | 1 | 52 | 0.836 | 0 | 1.939 | 1.729 | 2.175 | 2.81107 | 2.63766 | 2.98448 | 1.419 | -0.718 |
| Log10 Stem Length (cm) | Log10 Stem Diameter (cm) | Subtree data | E. diversicolor | 2 | 57 | 0.781 | 0 | 2.155 | 1.9 | 2.445 | 2.84616 | 2.66482 | 3.0275 | 1.505 | -0.622 |
| Log10 Stem Length (cm) | Log10 Stem Diameter (cm) | Subtree data | E. diversicolor | 3 | 59 | 0.86 | 0 | 2.126 | 1.925 | 2.347 | 2.775 | 2.62925 | 2.92075 | 1.419 | -0.638 |
| Log10 Stem Length (cm) | Log10 Stem Diameter (cm) | Subtree data | E. diversicolor | 4 | 35 | 0.936 | 0 | 1.712 | 1.565 | 1.873 | 2.65041 | 2.54549 | 2.75534 | 1.613 | -0.606 |
| Log10 Stem Length (cm) | Log10 Stem Diameter (cm) | Subtree data | E. incrassata | 1 | 43 | 0.887 | 0 | 1.93 | 1.736 | 2.145 | 2.6966 | 2.54364 | 2.84957 | 1.356 | -0.695 |
| Log10 Stem Length (cm) | Log10 Stem Diameter (cm) | Subtree data | E. incrassata | 2 | 25 | 0.625 | 0 | 2.698 | 2.078 | 3.504 | 3.62587 | 3.00517 | 4.24657 | 1.357 | -0.841 |
| Log10 Stem Length (cm) | Log10 Stem Diameter (cm) | Subtree data | E. incrassata | 3 | 15 | 0.786 | 0 | 2.487 | 1.891 | 3.27 | 3.31465 | 2.77221 | 3.85708 | 1.401 | -0.77 |
| Log10 Stem Surface Area | Log10 Stem Volume (cm3) | Raw data | E. gomphocephela | 1 | 36 | 0.892 | 0 | 0.6501 | 0.5799 | 0.7289 | 1.031 | 0.946 | 1.115 | 0.443 | -0.904 |
| Log10 Stem Surface Area | Log10 Stem Volume (cm3) | Raw data | E. gomphocephela | 2 | 43 | 0.651 | 0 | 0.7422 | 0.6166 | 0.8933 | 1.074 | 0.986 | 1.163 | 0.761 | -0.422 |
| Log10 Stem Surface Area | Log10 Stem Volume (cm3) | Raw data | E. gomphocephela | 3 | 41 | 0.673 | 0 | 0.799 | 0.6647 | 0.9605 | 1.09 | 1.003 | 1.176 | 0.822 | -0.335 |
| Log10 Stem Surface Area | Log10 Stem Volume (cm3) | Raw data | E. gomphocephela | 4 | 33 | 0.832 | 0 | 0.8609 | 0.7414 | 0.9998 | 1.158 | 1.037 | 1.279 | 0.558 | -0.697 |
| Log10 Stem Surface Area | Log10 Stem Volume (cm3) | Raw data | E. gomphocephela | 5 | 25 | 0.768 | 0 | 0.7184 | 0.5845 | 0.8828 | 1.072 | 0.972 | 1.171 | 0.796 | -0.384 |
| Log10 Stem Surface Area | Log10 Stem Volume (cm3) | Raw data | E. gomphocephela | 6 | 25 | 0.817 | 0 | 0.9546 | 0.7946 | 1.1468 | 1.23 | 1.083 | 1.376 | 0.633 | -0.626 |
| Log10 Stem Surface Area | Log10 Stem Volume (cm3) | Raw data | E. ceasia | 1 | 13 | 0.98 | 0 | 0.7163 | 0.6527 | 0.7862 | 1.084 | 0.988 | 1.18 | 0.327 | -1.057 |
| Log10 Stem Surface Area | Log10 Stem Volume (cm3) | Raw data | E. ceasia | 2 | 17 | 0.985 | 0 | 0.7482 | 0.6995 | 0.8003 | 1.107 | 1.002 | 1.213 | -0.189 | -1.733 |
| Log10 Stem Surface Area | Log10 Stem Volume (cm3) | Raw data | E. ceasia | 3 | 56 | 0.958 | 0 | 0.7092 | 0.6707 | 0.7499 | 1.053 | 0.989 | 1.117 | 0.153 | -1.269 |
| Log10 Stem Surface Area | Log10 Stem Volume (cm3) | Raw data | E. diversicolor | 1 | 52 | 0.836 | 0 | 0.7394 | 0.6591 | 0.8293 | 1.129 | 1.044 | 1.214 | 0.588 | -0.732 |
| Log10 Stem Surface Area | Log10 Stem Volume (cm3) | Raw data | E. diversicolor | 2 | 57 | 0.816 | 0 | 0.8778 | 0.7819 | 0.9855 | 1.159 | 1.081 | 1.237 | 0.691 | -0.534 |
| Log10 Stem Surface Area | Log10 Stem Volume (cm3) | Raw data | E. diversicolor | 3 | 59 | 0.813 | 0 | 0.7125 | 0.6354 | 0.799 | 1.093 | 1.03 | 1.155 | 0.728 | -0.512 |
| Log10 Stem Surface Area | Log10 Stem Volume (cm3) | Raw data | E. diversicolor | 4 | 35 | 0.666 | 0 | 0.6483 | 0.529 | 0.7944 | 1.112 | 1.031 | 1.193 | 0.934 | -0.274 |
| Log10 Stem Surface Area | Log10 Stem Volume (cm3) | Raw data | E. incrassata | 1 | 43 | 0.895 | 0 | 0.6937 | 0.6265 | 0.7681 | 1.086 | 1.018 | 1.155 | 0.609 | -0.687 |
| Log10 Stem Surface Area | Log10 Stem Volume (cm3) | Raw data | E. incrassata | 2 | 25 | 0.812 | 0 | 0.8318 | 0.6905 | 1.0019 | 1.268 | 1.089 | 1.448 | 0.406 | -1.037 |
| Log10 Stem Surface Area | Log10 Stem Volume (cm3) | Raw data | E. incrassata | 3 | 15 | 0.906 | 0 | 0.8463 | 0.705 | 1.016 | 1.245 | 1.094 | 1.395 | 0.544 | -0.828 |
| Log10 Stem Surface Area | Log10 Stem Volume (cm3) | Subtree data | E. gomphocephela | 1 | 36 | 0.996 | 0 | 0.7872 | 0.7709 | 0.8039 | 1.31 | 1.292 | 1.328 | 0.824 | -0.616 |
| Log10 Stem Surface Area | Log10 Stem Volume (cm3) | Subtree data | E. gomphocephela | 2 | 43 | 0.993 | 0 | 0.8193 | 0.7975 | 0.8417 | 1.305 | 1.288 | 1.322 | 1.317 | 0.014 |
| Log10 Stem Surface Area | Log10 Stem Volume (cm3) | Subtree data | E. gomphocephela | 3 | 41 | 0.992 | 0 | 0.8349 | 0.8115 | 0.8589 | 1.298 | 1.28 | 1.315 | 1.376 | 0.094 |
| Log10 Stem Surface Area | Log10 Stem Volume (cm3) | Subtree data | E. gomphocephela | 4 | 33 | 0.991 | 0 | 0.81 | 0.7821 | 0.839 | 1.334 | 1.314 | 1.354 | 1.16 | -0.215 |
| Log10 Stem Surface Area | Log10 Stem Volume (cm3) | Subtree data | E. gomphocephela | 5 | 25 | 0.994 | 0 | 0.813 | 0.7858 | 0.8411 | 1.303 | 1.282 | 1.323 | 1.297 | -0.007 |
| Log10 Stem Surface Area | Log10 Stem Volume (cm3) | Subtree data | E. gomphocephela | 6 | 25 | 0.98 | 0 | 0.8287 | 0.7797 | 0.8808 | 1.361 | 1.33 | 1.393 | 1.236 | -0.151 |
| Log10 Stem Surface Area | Log10 Stem Volume (cm3) | Subtree data | E. ceasia | 1 | 13 | 0.998 | 0 | 0.769 | 0.746 | 0.7927 | 1.231 | 1.198 | 1.264 | 0.575 | -0.853 |
| Log10 Stem Surface Area | Log10 Stem Volume (cm3) | Subtree data | E. ceasia | 2 | 17 | 0.998 | 0 | 0.7721 | 0.7542 | 0.7904 | 1.238 | 1.203 | 1.274 | 0.059 | -1.526 |
| Log10 Stem Surface Area | Log10 Stem Volume (cm3) | Subtree data | E. ceasia | 3 | 56 | 0.991 | 0 | 0.7433 | 0.7243 | 0.7629 | 1.207 | 1.177 | 1.237 | 0.453 | -1.014 |
| Log10 Stem Surface Area | Log10 Stem Volume (cm3) | Subtree data | E. diversicolor | 1 | 52 | 0.991 | 0 | 0.7999 | 0.7783 | 0.822 | 1.35 | 1.33 | 1.37 | 1.054 | -0.369 |
| Log10 Stem Surface Area | Log10 Stem Volume (cm3) | Subtree data | E. diversicolor | 2 | 57 | 0.987 | 0 | 0.8412 | 0.8156 | 0.8677 | 1.312 | 1.295 | 1.329 | 1.255 | -0.068 |
| Log10 Stem Surface Area | Log10 Stem Volume (cm3) | Subtree data | E. diversicolor | 3 | 59 | 0.987 | 0 | 0.8271 | 0.8027 | 0.8522 | 1.303 | 1.283 | 1.323 | 1.163 | -0.169 |
| Log10 Stem Surface Area | Log10 Stem Volume (cm3) | Subtree data | E. diversicolor | 4 | 35 | 0.992 | 0 | 0.782 | 0.7584 | 0.8064 | 1.32 | 1.3 | 1.34 | 1.363 | 0.055 |
| Log10 Stem Surface Area | Log10 Stem Volume (cm3) | Subtree data | E. incrassata | 1 | 43 | 0.992 | 0 | 0.7848 | 0.7636 | 0.8066 | 1.309 | 1.289 | 1.328 | 1.035 | -0.348 |
| Log10 Stem Surface Area | Log10 Stem Volume (cm3) | Subtree data | E. incrassata | 2 | 25 | 0.987 | 0 | 0.8777 | 0.8351 | 0.9225 | 1.453 | 1.412 | 1.495 | 0.908 | -0.622 |
| Log10 Stem Surface Area | Log10 Stem Volume (cm3) | Subtree data | E. incrassata | 3 | 15 | 0.982 | 0 | 0.8443 | 0.7787 | 0.9155 | 1.378 | 1.333 | 1.424 | 1.043 | -0.397 |
| Log10 Stem Diameter (cm) | Log10 Stem Volume (cm3) | Raw data | E. gomphocephela | 1 | 36 | 0.765 | 0 | 0.4413 | 0.373 | 0.5221 | -0.3461 | -0.4311 | -0.2611 | -0.745 | -0.904 |
| Log10 Stem Diameter (cm) | Log10 Stem Volume (cm3) | Raw data | E. gomphocephela | 2 | 43 | 0.456 | 0 | 0.5944 | 0.472 | 0.7486 | -0.3302 | -0.4202 | -0.2402 | -0.581 | -0.422 |
| Log10 Stem Diameter (cm) | Log10 Stem Volume (cm3) | Raw data | E. gomphocephela | 3 | 41 | 0.363 | 0 | 0.572 | 0.4429 | 0.7387 | -0.3633 | -0.4539 | -0.2727 | -0.555 | -0.335 |
| Log10 Stem Diameter (cm) | Log10 Stem Volume (cm3) | Raw data | E. gomphocephela | 4 | 33 | 0.27 | 0.002 | 0.4129 | 0.3034 | 0.5618 | -0.3649 | -0.4926 | -0.2372 | -0.653 | -0.697 |
| Log10 Stem Diameter (cm) | Log10 Stem Volume (cm3) | Raw data | E. gomphocephela | 5 | 25 | 0.534 | 0 | 0.5066 | 0.379 | 0.6773 | -0.3834 | -0.4857 | -0.2811 | -0.578 | -0.384 |
| Log10 Stem Diameter (cm) | Log10 Stem Volume (cm3) | Raw data | E. gomphocephela | 6 | 25 | 0.101 | 0.121 | 0.4306 | 0.2892 | 0.6413 | -0.3868 | -0.5468 | -0.2269 | -0.656 | -0.626 |
| Log10 Stem Diameter (cm) | Log10 Stem Volume (cm3) | Raw data | E. ceasia | 1 | 13 | 0.893 | 0 | 0.3077 | 0.2481 | 0.3816 | -0.4566 | -0.5532 | -0.3599 | -0.782 | -1.057 |
| Log10 Stem Diameter (cm) | Log10 Stem Volume (cm3) | Raw data | E. ceasia | 2 | 17 | 0.888 | 0 | 0.2732 | 0.2274 | 0.3282 | -0.4679 | -0.5739 | -0.362 | -0.941 | -1.733 |
| Log10 Stem Diameter (cm) | Log10 Stem Volume (cm3) | Raw data | E. ceasia | 3 | 56 | 0.816 | 0 | 0.3386 | 0.3013 | 0.3805 | -0.3902 | -0.4545 | -0.326 | -0.82 | -1.269 |
| Log10 Stem Diameter (cm) | Log10 Stem Volume (cm3) | Raw data | E. diversicolor | 1 | 52 | 0.539 | 0 | 0.4414 | 0.3644 | 0.5346 | -0.3948 | -0.4814 | -0.3082 | -0.718 | -0.732 |
| Log10 Stem Diameter (cm) | Log10 Stem Volume (cm3) | Raw data | E. diversicolor | 2 | 57 | 0.232 | 0 | 0.4299 | 0.34 | 0.5436 | -0.3927 | -0.4758 | -0.3097 | -0.622 | -0.534 |
| Log10 Stem Diameter (cm) | Log10 Stem Volume (cm3) | Raw data | E. diversicolor | 3 | 59 | 0.574 | 0 | 0.4723 | 0.3976 | 0.5612 | -0.396 | -0.4598 | -0.3323 | -0.638 | -0.512 |
| Log10 Stem Diameter (cm) | Log10 Stem Volume (cm3) | Raw data | E. diversicolor | 4 | 35 | 0.612 | 0 | 0.6017 | 0.4835 | 0.7488 | -0.441 | -0.5225 | -0.3594 | -0.606 | -0.274 |
| Log10 Stem Diameter (cm) | Log10 Stem Volume (cm3) | Raw data | E. incrassata | 1 | 43 | 0.701 | 0 | 0.4105 | 0.3457 | 0.4873 | -0.4125 | -0.4821 | -0.343 | -0.695 | -0.687 |
| Log10 Stem Diameter (cm) | Log10 Stem Volume (cm3) | Raw data | E. incrassata | 2 | 25 | 0.325 | 0.003 | 0.4394 | 0.3105 | 0.6219 | -0.3851 | -0.5677 | -0.2025 | -0.841 | -1.037 |
| Log10 Stem Diameter (cm) | Log10 Stem Volume (cm3) | Raw data | E. incrassata | 3 | 15 | 0.36 | 0.018 | 0.3243 | 0.2041 | 0.5151 | -0.5012 | -0.6558 | -0.3467 | -0.77 | -0.828 |
| Log10 Stem Diameter (cm) | Log10 Stem Volume (cm3) | Subtree data | E. gomphocephela | 1 | 36 | 0.972 | 0 | 0.3326 | 0.3137 | 0.3526 | -0.5398 | -0.561 | -0.5187 | -0.745 | -0.616 |
| Log10 Stem Diameter (cm) | Log10 Stem Volume (cm3) | Subtree data | E. gomphocephela | 2 | 43 | 0.972 | 0 | 0.3548 | 0.3365 | 0.374 | -0.5865 | -0.6008 | -0.5722 | -0.581 | 0.014 |
| Log10 Stem Diameter (cm) | Log10 Stem Volume (cm3) | Subtree data | E. gomphocephela | 3 | 41 | 0.957 | 0 | 0.3646 | 0.3408 | 0.3901 | -0.5895 | -0.6074 | -0.5715 | -0.555 | 0.094 |
| Log10 Stem Diameter (cm) | Log10 Stem Volume (cm3) | Subtree data | E. gomphocephela | 4 | 33 | 0.974 | 0 | 0.3751 | 0.3537 | 0.3977 | -0.572 | -0.5876 | -0.5564 | -0.653 | -0.215 |
| Log10 Stem Diameter (cm) | Log10 Stem Volume (cm3) | Subtree data | E. gomphocephela | 5 | 25 | 0.978 | 0 | 0.3623 | 0.34 | 0.386 | -0.5754 | -0.5924 | -0.5583 | -0.578 | -0.007 |
| Log10 Stem Diameter (cm) | Log10 Stem Volume (cm3) | Subtree data | E. gomphocephela | 6 | 25 | 0.981 | 0 | 0.3829 | 0.3608 | 0.4064 | -0.5983 | -0.6125 | -0.5841 | -0.656 | -0.151 |
| Log10 Stem Diameter (cm) | Log10 Stem Volume (cm3) | Subtree data | E. ceasia | 1 | 13 | 0.979 | 0 | 0.2723 | 0.2475 | 0.2996 | -0.5495 | -0.5858 | -0.5132 | -0.782 | -0.853 |
| Log10 Stem Diameter (cm) | Log10 Stem Volume (cm3) | Subtree data | E. ceasia | 2 | 17 | 0.973 | 0 | 0.2609 | 0.2385 | 0.2854 | -0.5431 | -0.5891 | -0.4971 | -0.941 | -1.526 |
| Log10 Stem Diameter (cm) | Log10 Stem Volume (cm3) | Subtree data | E. ceasia | 3 | 56 | 0.951 | 0 | 0.2866 | 0.2698 | 0.3044 | -0.5294 | -0.5562 | -0.5026 | -0.82 | -1.014 |
| Log10 Stem Diameter (cm) | Log10 Stem Volume (cm3) | Subtree data | E. diversicolor | 1 | 52 | 0.914 | 0 | 0.3471 | 0.3194 | 0.3771 | -0.5896 | -0.6162 | -0.5629 | -0.718 | -0.369 |
| Log10 Stem Diameter (cm) | Log10 Stem Volume (cm3) | Subtree data | E. diversicolor | 2 | 57 | 0.907 | 0 | 0.3463 | 0.3189 | 0.376 | -0.5986 | -0.6177 | -0.5794 | -0.622 | -0.068 |
| Log10 Stem Diameter (cm) | Log10 Stem Volume (cm3) | Subtree data | E. diversicolor | 3 | 59 | 0.954 | 0 | 0.3325 | 0.3142 | 0.3519 | -0.5819 | -0.5971 | -0.5668 | -0.638 | -0.169 |
| Log10 Stem Diameter (cm) | Log10 Stem Volume (cm3) | Subtree data | E. diversicolor | 4 | 35 | 0.983 | 0 | 0.3732 | 0.3563 | 0.3909 | -0.6266 | -0.6412 | -0.6121 | -0.606 | 0.055 |
| Log10 Stem Diameter (cm) | Log10 Stem Volume (cm3) | Subtree data | E. incrassata | 1 | 43 | 0.949 | 0 | 0.3334 | 0.3104 | 0.3581 | -0.5785 | -0.6001 | -0.557 | -0.695 | -0.348 |
| Log10 Stem Diameter (cm) | Log10 Stem Volume (cm3) | Subtree data | E. incrassata | 2 | 25 | 0.778 | 0 | 0.2965 | 0.2423 | 0.3629 | -0.6564 | -0.7148 | -0.598 | -0.841 | -0.622 |

**Table S3.** SMA regression data for the 4 species within the Eucalyptus genus (each comprised of multiple individuals as described in the Methods). Data are grouped by relationship (first two columns) and whether the regressions were done on raw data or subtrees as described in the methods.

| Y variable | X variable | Data | Species | n | R2 | p | Slope | LowCI | UppCI | Interc | LowCI | UppCI | Ymean | Xmean |
|---|---|---|---|---|---|---|---|---|---|---|---|---|---|---|
| Log10 Stem Length (cm) | Log10 Stem Diameter (cm) | Raw | E. gomphocephela | 203 | 0.117 | 0 | -1.675 | -1.908 | -1.47 | -0.2468 | -0.4024 | -0.0911 | 0.801 | -0.625 |
| Log10 Stem Length (cm) | Log10 Stem Diameter (cm) | Raw | E. ceasia | 94 | 0.391 | 0 | 1.582 | 1.347 | 1.858 | 1.7901 | 1.5567 | 2.0235 | 0.479 | -0.829 |
| Log10 Stem Length (cm) | Log10 Stem Diameter (cm) | Raw | E. diversicolor | 203 | 0.096 | 0 | -1.635 | -1.866 | -1.433 | -0.1921 | -0.3504 | -0.0338 | 0.868 | -0.648 |
| Log10 Stem Length (cm) | Log10 Stem Diameter (cm) | Raw | E. incrassata | 88 | 0.001 | 0.8 | -1.566 | -1.937 | -1.266 | -0.3349 | -0.6104 | -0.0593 | 0.816 | -0.735 |
| Log10 Stem Length (cm) | Log10 Stem Diameter (cm) | Subtrees | E. gomphocephela | 203 | 0.918 | 0 | 1.934 | 1.858 | 2.013 | 2.6934 | 2.6405 | 2.7462 | 1.484 | -0.625 |
| Log10 Stem Length (cm) | Log10 Stem Diameter (cm) | Subtrees | E. ceasia | 94 | 0.901 | 0 | 1.978 | 1.853 | 2.111 | 2.471 | 2.3549 | 2.587 | 0.831 | -0.829 |
| Log10 Stem Length (cm) | Log10 Stem Diameter (cm) | Subtrees | E. diversicolor | 203 | 0.835 | 0 | 1.981 | 1.872 | 2.096 | 2.7614 | 2.6822 | 2.8406 | 1.477 | -0.648 |
| Log10 Stem Length (cm) | Log10 Stem Diameter (cm) | Subtrees | E. incrassata | 88 | 0.701 | 0 | 1.996 | 1.775 | 2.244 | 2.8463 | 2.6628 | 3.0299 | 1.38 | -0.735 |
| Log10 Stem Surface Area (cm2) | Log10 Stem Volume (cm2) | Raw | E. gomphocephela | 203 | 0.794 | 0 | 0.7676 | 0.7206 | 0.8176 | 1.098 | 1.059 | 1.138 | 0.672 | -0.555 |
| Log10 Stem Surface Area (cm2) | Log10 Stem Volume (cm2) | Raw | E. ceasia | 94 | 0.973 | 0 | 0.7239 | 0.6999 | 0.7487 | 1.076 | 1.035 | 1.118 | 0.147 | -1.284 |
| Log10 Stem Surface Area (cm2) | Log10 Stem Volume (cm2) | Raw | E. diversicolor | 203 | 0.804 | 0 | 0.7567 | 0.7116 | 0.8047 | 1.121 | 1.083 | 1.158 | 0.717 | -0.533 |
| Log10 Stem Surface Area (cm2) | Log10 Stem Volume (cm2) | Raw | E. incrassata | 88 | 0.884 | 0 | 0.7322 | 0.6807 | 0.7875 | 1.133 | 1.079 | 1.187 | 0.578 | -0.758 |
| Log10 Stem Surface Area (cm2) | Log10 Stem Volume (cm2) | Subtrees | E. gomphocephela | 203 | 0.992 | 0 | 0.8095 | 0.7993 | 0.8198 | 1.318 | 1.31 | 1.326 | 1.204 | -0.142 |
| Log10 Stem Surface Area (cm2) | Log10 Stem Volume (cm2) | Subtrees | E. ceasia | 94 | 0.994 | 0 | 0.7515 | 0.7399 | 0.7633 | 1.212 | 1.193 | 1.232 | 0.422 | -1.051 |
| Log10 Stem Surface Area (cm2) | Log10 Stem Volume (cm2) | Subtrees | E. diversicolor | 203 | 0.988 | 0 | 0.81 | 0.7977 | 0.8225 | 1.32 | 1.31 | 1.33 | 1.196 | -0.153 |
| Log10 Stem Surface Area (cm2) | Log10 Stem Volume (cm2) | Subtrees | E. incrassata | 88 | 0.98 | 0 | 0.7948 | 0.771 | 0.8193 | 1.34 | 1.318 | 1.361 | 1.033 | -0.386 |
| Log10 Stem Diameter (cm) | Log10 Stem Volume (cm2) | Raw | E. gomphocephela | 203 | 0.451 | 0 | 0.4708 | 0.4247 | 0.5218 | -0.3641 | -0.4054 | -0.3229 | -0.625 | -0.555 |
| Log10 Stem Diameter (cm) | Log10 Stem Volume (cm2) | Raw | E. ceasia | 94 | 0.854 | 0 | 0.3092 | 0.2857 | 0.3346 | -0.4319 | -0.4739 | -0.39 | -0.829 | -1.284 |
| Log10 Stem Diameter (cm) | Log10 Stem Volume (cm2) | Raw | E. diversicolor | 203 | 0.479 | 0 | 0.4641 | 0.4199 | 0.513 | -0.4009 | -0.4398 | -0.3619 | -0.648 | -0.533 |
| Log10 Stem Diameter (cm) | Log10 Stem Volume (cm2) | Raw | E. incrassata | 88 | 0.61 | 0 | 0.399 | 0.3491 | 0.456 | -0.4319 | -0.487 | -0.3768 | -0.735 | -0.758 |
| Log10 Stem Diameter (cm) | Log10 Stem Volume (cm2) | Subtrees | E. gomphocephela | 203 | 0.964 | 0 | 0.3502 | 0.3411 | 0.3596 | -0.5758 | -0.5833 | -0.5684 | -0.625 | -0.142 |
| Log10 Stem Diameter (cm) | Log10 Stem Volume (cm2) | Subtrees | E. ceasia | 94 | 0.964 | 0 | 0.2758 | 0.2652 | 0.2868 | -0.5391 | -0.5568 | -0.5214 | -0.829 | -1.051 |
| Log10 Stem Diameter (cm) | Log10 Stem Volume (cm2) | Subtrees | E. diversicolor | 203 | 0.936 | 0 | 0.3463 | 0.3343 | 0.3587 | -0.5954 | -0.6053 | -0.5855 | -0.648 | -0.153 |
| Log10 Stem Diameter (cm) | Log10 Stem Volume (cm2) | Subtrees | E. incrassata | 88 | 0.901 | 0 | 0.3322 | 0.3105 | 0.3553 | -0.6062 | -0.6261 | -0.5862 | -0.735 | -0.386 |
| Log10 Stem Length (cm) | Log10 Stem Volume (cm2) | Raw | E. gomphocephela | 203 | 0.217 | 0 | 0.7885 | 0.6974 | 0.8914 | 1.238 | 1.153 | 1.324 | 0.801 | -0.555 |
| Log10 Stem Length (cm) | Log10 Stem Volume (cm2) | Raw | E. ceasia | 94 | 0.767 | 0 | 0.4891 | 0.4427 | 0.5404 | 1.107 | 1.022 | 1.191 | 0.479 | -1.284 |
| Log10 Stem Length (cm) | Log10 Stem Volume (cm2) | Raw | E. diversicolor | 203 | 0.222 | 0 | 0.759 | 0.6715 | 0.8578 | 1.273 | 1.192 | 1.355 | 0.868 | -0.533 |
| Log10 Stem Length (cm) | Log10 Stem Volume (cm2) | Raw | E. incrassata | 88 | 0.364 | 0 | 0.6249 | 0.5271 | 0.7408 | 1.289 | 1.177 | 1.402 | 0.816 | -0.758 |
| Log10 Stem Length (cm) | Log10 Stem Volume (cm2) | Subtrees | E. gomphocephela | 203 | 0.943 | 0 | 0.6772 | 0.6551 | 0.7002 | 1.58 | 1.561 | 1.598 | 1.484 | -0.142 |
| Log10 Stem Length (cm) | Log10 Stem Volume (cm2) | Subtrees | E. ceasia | 94 | 0.966 | 0 | 0.5456 | 0.5251 | 0.5669 | 1.405 | 1.37 | 1.439 | 0.831 | -1.051 |
| Log10 Stem Length (cm) | Log10 Stem Volume (cm2) | Subtrees | E. diversicolor | 203 | 0.919 | 0 | 0.6861 | 0.6595 | 0.7138 | 1.582 | 1.56 | 1.604 | 1.477 | -0.153 |
| Log10 Stem Length (cm) | Log10 Stem Volume (cm2) | Subtrees | E. incrassata | 88 | 0.883 | 0 | 0.6629 | 0.616 | 0.7133 | 1.637 | 1.593 | 1.68 | 1.38 | -0.386 |
| Log10 Stem Diameter (cm) | Log10 Stem Surface Area (cm2) | Raw | E. gomphocephela | 203 | 0.069 | 0 | 0.6133 | 0.5365 | 0.7011 | -1.0378 | -1.1104 | -0.9653 | -0.625 | 0.672 |
| Log10 Stem Diameter (cm) | Log10 Stem Surface Area (cm2) | Raw | E. ceasia | 94 | 0.722 | 0 | 0.4272 | 0.3831 | 0.4763 | -0.8918 | -0.9316 | -0.852 | -0.829 | 0.147 |
| Log10 Stem Diameter (cm) | Log10 Stem Surface Area (cm2) | Raw | E. diversicolor | 203 | 0.091 | 0 | 0.6133 | 0.5373 | 0.7 | -1.0882 | -1.162 | -1.0144 | -0.648 | 0.717 |
| Log10 Stem Diameter (cm) | Log10 Stem Surface Area (cm2) | Raw | E. incrassata | 88 | 0.272 | 0 | 0.545 | 0.4543 | 0.6537 | -1.0496 | -1.1293 | -0.9698 | -0.735 | 0.578 |
| Log10 Stem Diameter (cm) | Log10 Stem Surface Area (cm2) | Subtrees | E. gomphocephela | 203 | 0.955 | 0 | 0.4326 | 0.4201 | 0.4456 | -1.1461 | -1.1635 | -1.1288 | -0.625 | 1.204 |
| Log10 Stem Diameter (cm) | Log10 Stem Surface Area (cm2) | Subtrees | E. ceasia | 94 | 0.942 | 0 | 0.367 | 0.3492 | 0.3858 | -0.984 | -1.0031 | -0.965 | -0.829 | 0.422 |
| Log10 Stem Diameter (cm) | Log10 Stem Surface Area (cm2) | Subtrees | E. diversicolor | 203 | 0.909 | 0 | 0.4275 | 0.41 | 0.4458 | -1.1596 | -1.184 | -1.1352 | -0.648 | 1.196 |
| Log10 Stem Diameter (cm) | Log10 Stem Surface Area (cm2) | Subtrees | E. incrassata | 88 | 0.834 | 0 | 0.4179 | 0.383 | 0.456 | -1.1661 | -1.2105 | -1.1217 | -0.735 | 1.033 |
| Log10 Stem Length (cm) | Log10 Stem Surface Area (cm2) | Raw | E. gomphocephela | 203 | 0.668 | 0 | 1.0272 | 0.9481 | 1.1128 | 0.11 | 0.0423 | 0.1778 | 0.801 | 0.672 |
| Log10 Stem Length (cm) | Log10 Stem Surface Area (cm2) | Raw | E. ceasia | 94 | 0.889 | 0 | 0.6757 | 0.6306 | 0.7239 | 0.3795 | 0.3406 | 0.4184 | 0.479 | 0.147 |
| Log10 Stem Length (cm) | Log10 Stem Surface Area (cm2) | Raw | E. diversicolor | 203 | 0.66 | 0 | 1.003 | 0.925 | 1.0876 | 0.1491 | 0.0794 | 0.2189 | 0.868 | 0.717 |
| Log10 Stem Length (cm) | Log10 Stem Surface Area (cm2) | Raw | E. incrassata | 88 | 0.703 | 0 | 0.8535 | 0.7596 | 0.959 | 0.322 | 0.2457 | 0.3984 | 0.816 | 0.578 |
| Log10 Stem Length (cm) | Log10 Stem Surface Area (cm2) | Subtrees | E. gomphocephela | 203 | 0.977 | 0 | 0.8367 | 0.819 | 0.8547 | 0.4768 | 0.4525 | 0.5012 | 1.484 | 1.204 |
| Log10 Stem Length (cm) | Log10 Stem Surface Area (cm2) | Subtrees | E. ceasia | 94 | 0.986 | 0 | 0.726 | 0.7085 | 0.744 | 0.5244 | 0.506 | 0.5427 | 0.831 | 0.422 |
| Log10 Stem Length (cm) | Log10 Stem Surface Area (cm2) | Subtrees | E. diversicolor | 203 | 0.968 | 0 | 0.847 | 0.8261 | 0.8684 | 0.464 | 0.4353 | 0.4928 | 1.477 | 1.196 |
| Log10 Stem Length (cm) | Log10 Stem Surface Area (cm2) | Subtrees | E. incrassata | 88 | 0.956 | 0 | 0.834 | 0.7974 | 0.8723 | 0.5191 | 0.4738 | 0.5645 | 1.38 | 1.033 |

**Table S4.** Regression statistics for the bivariate relationships depected in Fig. 4 of the main manuscript.

| Y variable | X variable | n | R2 | p | Slope | LowCI | UppCI | Interc | LowCI | UppCI | Ymean | Xmean |
|---|---|---|---|---|---|---|---|---|---|---|---|---|
| Log Height (m) | Log Diameter (m) | 1387 | 0.688 | 0 | 1.102 | 1.07 | 1.135 | 1.827 | 1.751 | 1.903 | -0.681 | -2.276 |
| Log Height (m) | Log Mass (g) | 1225 | 0.765 | 0 | 0.4347 | 0.423 | 0.4467 | 0.08435 | 0.05723 | 0.11146 | -0.652 | -1.694 |
| Log Diameter (m) | Log Mass (g) | 1290 | 0.905 | 0 | 0.4176 | 0.4107 | 0.4247 | -1.584 | -1.601 | -1.567 | -2.358 | -1.852 |

**Table S5.** SMA regression statistics for bivariate relationships between vein length, diameter, surface area and volume, for the three leaves described in the Methods.

| Y variable | X variable | Species | n | R2 | p | Slope | LowCI | UppCI | Interc | LowCI | UppCI | Ymean | Xmean |
|---|---|---|---|---|---|---|---|---|---|---|---|---|---|
| Log10 Length (mm) | Log10 Diameter (mm) | B. victoriae | 9502 | 0.011 | 0 | 1.939 | 1.901 | 1.979 | 1.73 | 1.678 | 1.782 | -0.822 | -1.316 |
| Log10 Length (mm) | Log10 Diameter (mm) | H. comptoniana | 24148 | 0.182 | 0 | 1.958 | 1.935 | 1.98 | 2.003 | 1.97 | 2.037 | -0.913 | -1.49 |
| Log10 Length (mm) | Log10 Diameter (mm) | L. cuneata | 3096 | 0.091 | 0 | 2.684 | 2.595 | 2.775 | 4.097 | 3.922 | 4.272 | -1.089 | -1.933 |
| Log10 Surface Area (mm2) | Log10 Volume (mm3) | B. victoriae | 9502 | 0.916 | 0 | 0.7765 | 0.7719 | 0.781 | 0.8887 | 0.8738 | 0.9037 | -1.641 | -3.258 |
| Log10 Surface Area (mm2) | Log10 Volume (mm3) | H. comptoniana | 24148 | 0.957 | 0 | 0.7629 | 0.7609 | 0.7649 | 0.9143 | 0.9068 | 0.9218 | -1.905 | -3.696 |
| Log10 Surface Area (mm2) | Log10 Volume (mm3) | L. cuneata | 3096 | 0.954 | 0 | 0.8247 | 0.8184 | 0.8309 | 1.3993 | 1.3694 | 1.4293 | -2.524 | -4.758 |
| Log10 Diameter (mm) | Log10 Volume (mm3) | B. victoriae | 9502 | 0.565 | 0 | 0.3418 | 0.3373 | 0.3463 | -0.2025 | -0.2175 | -0.1875 | -1.316 | -3.258 |
| Log10 Diameter (mm) | Log10 Volume (mm3) | H. comptoniana | 24148 | 0.72 | 0 | 0.2991 | 0.2972 | 0.3012 | -0.384 | -0.3916 | -0.3765 | -1.49 | -3.696 |
| Log10 Diameter (mm) | Log10 Volume (mm3) | L. cuneata | 3096 | 0.547 | 0 | 0.2631 | 0.257 | 0.2694 | -0.6806 | -0.7106 | -0.6506 | -1.933 | -4.758 |
| Log10 Length (mm) | Log10 Volume (mm3) | B. victoriae | 9502 | 0.538 | 0 | 0.6628 | 0.6538 | 0.672 | 1.337 | 1.308 | 1.367 | -0.822 | -3.258 |
| Log10 Length (mm) | Log10 Volume (mm3) | H. comptoniana | 24148 | 0.707 | 0 | 0.5856 | 0.5816 | 0.5896 | 1.252 | 1.237 | 1.267 | -0.913 | -3.696 |
| Log10 Length (mm) | Log10 Volume (mm3) | L. cuneata | 3096 | 0.748 | 0 | 0.7061 | 0.6938 | 0.7187 | 2.271 | 2.211 | 2.331 | -1.089 | -4.758 |
| Log10 Diameter (mm) | Log10 Surface Area (mm2) | B. victoriae | 9502 | 0.279 | 0 | 0.4402 | 0.4327 | 0.4477 | -0.5937 | -0.6065 | -0.5809 | -1.316 | -1.641 |
| Log10 Diameter (mm) | Log10 Surface Area (mm2) | H. comptoniana | 24148 | 0.518 | 0 | 0.3921 | 0.3887 | 0.3956 | -0.7426 | -0.7494 | -0.7357 | -1.49 | -1.905 |
| Log10 Diameter (mm) | Log10 Surface Area (mm2) | L. cuneata | 3096 | 0.334 | 0 | 0.3191 | 0.31 | 0.3284 | -1.1271 | -1.1509 | -1.1033 | -1.933 | -2.524 |
| Log10 Length (mm) | Log10 Surface Area (mm2) | B. victoriae | 9502 | 0.808 | 0 | 0.8537 | 0.8462 | 0.8612 | 0.5788 | 0.5661 | 0.5915 | -0.822 | -1.641 |
| Log10 Length (mm) | Log10 Surface Area (mm2) | H. comptoniana | 24148 | 0.874 | 0 | 0.7676 | 0.7642 | 0.7711 | 0.5498 | 0.543 | 0.5566 | -0.913 | -1.905 |
| Log10 Length (mm) | Log10 Surface Area (mm2) | L. cuneata | 3096 | 0.907 | 0 | 0.8563 | 0.8471 | 0.8655 | 1.0725 | 1.0489 | 1.0962 | -1.089 | -2.524 |